%% file: paper.tex
\documentclass{article}

\input{packages}

\usepackage[preprint]{mlsys2025}
\usepackage{enumitem}
\input{marcos}

\mlsystitlerunning{\sys{}}
\hypersetup{pdfsubject={XM-Sparse Preprint}}

\newif\ifdraft \draftfalse

\ifdraft
  \newcommand{\xd}[1]{{\color{blue}[XD: #1]}}
  \newcommand{\cl}[1]{{\color{orange}[Claude: #1]}}
  \newcommand{\reviewcomment}[1]{{\color{violet}[Reviewer: #1]}}
  \newcommand{\editadd}[1]{{\color{teal}#1}}
  \newcommand{\editdel}[1]{{\color{red}\sout{#1}}}
  
\else
  \newcommand{\xd}[1]{}
  \newcommand{\cl}[1]{}
  \newcommand{\reviewcomment}[1]{}
  \newcommand{\editadd}[1]{#1}
  \newcommand{\editdel}[1]{}
  
\fi

\begin{document}

\twocolumn[
\mlsystitle{\sys{}: High-Throughput Sparse-Attention LLM Serving\texorpdfstring{\\}{ }over Scale-up Memory Pools}

\input{sections/author_info}
\mlsyskeywords{Machine Learning, Systems}

\vskip 0.3in
\begin{abstract}
\input{sections/abstract}
\end{abstract}
]
\printAffiliationsAndNotice{\textsuperscript{\dag}Work done during an internship at Netpreme. }

\input{sections/introduction}
\input{sections/background}
\input{sections/analysis}
\input{sections/design}
\input{sections/evaluation}
\input{sections/related}
\input{sections/conclusion}

\bibliographystyle{mlsys2025}
\bibliography{reference}

\clearpage
\appendix
\input{sections/appendix}

\end{document}

%% file: packages.tex
\usepackage{microtype}
\usepackage{graphicx}
\usepackage{amsmath,amssymb,amsthm}
\usepackage{booktabs,tabularx,multirow,makecell}
\usepackage{subcaption}
\usepackage{xcolor}
\usepackage{xspace}
\usepackage{tikz}
\usepackage{comment}
\usepackage[normalem]{ulem}
\usepackage[hyphens]{url}
\usepackage{hyperref}

\graphicspath{{figures/}}

%% file: marcos.tex
\newcommand{\sys}{%
  \texorpdfstring{%
    \leavevmode
    \ifdefined\editadd
      XM-Sparse%
    \else
      XM-Sparse%
    \fi
  }{XM-Sparse}\xspace}

\newcommand{\secref}[1]{\hyperref[#1]{\S\ref*{#1}}}
\newcommand{\figref}[1]{\hyperref[#1]{Fig.~\ref*{#1}}}
\newcommand{\tabref}[1]{\hyperref[#1]{Table~\ref*{#1}}}
\newcommand{\algoref}[1]{\hyperref[#1]{Algorithm~\ref*{#1}}}

\newcommand{\parab}[1]{\paragraph{#1}}

\theoremstyle{definition}

%% file: sections/author_info.tex
\mlsyssetsymbol{intern}{\dag}
\begin{mlsysauthorlist}
\mlsysauthor{Yijun Sun}{netpreme,hkust,intern}
\mlsysauthor{Xudong Liao}{netpreme}
\mlsysauthor{Zhuobin Huang}{netpreme,nus,intern}
\mlsysauthor{Kai Chen}{hkust}
\mlsysauthor{Nikita Lazarev}{netpreme}
\mlsysauthor{Zhizhen Zhong}{netpreme}
\end{mlsysauthorlist}
\mlsysaffiliation{netpreme}{Netpreme}
\mlsysaffiliation{hkust}{The Hong Kong University of Science and Technology}
\mlsysaffiliation{nus}{National University of Singapore}

%% file: sections/abstract.tex
Sparse attention has been increasingly adopted by production LLMs for long-context inference, yet serving systems must still retain the full KV history, making GPU HBM capacity a key limit on serving concurrency. Existing systems offload the KV cache to host memory, but suffer from the limited bandwidth of PCIe and the limited capacity of host memory. Emerging scale-up memory pools (e.g., X-Mem) relieve both limits and open new opportunities for sparse-attention offloading. In this paper, we first conduct an in-depth analysis and reveal that, even with scale-up memory pools, decode throughput remains bounded by HBM capacity, as hot KV buffers and the fully resident indexer key (K-index) cache occupy substantial HBM. Moreover, the two caches exhibit distinct access patterns. Based on these insights, we present \sys{}, the first sparse-attention offloading system over scale-up memory expansion. The core idea of \sys{} is to offload both caches aggressively while matching each with an I/O mechanism tailored to its access pattern. Specifically, for dynamically selected KV entries, \sys{} fuses TMA-based retrieval with attention in a tile-level pipeline to hide on-demand retrieval latency. For predictable K-index accesses, it prefetches pages via SM-efficient GPU-initiated DMA and coordinates prefetching with latency-critical operations to avoid interfering with decoding. We have implemented \sys{} atop SGLang, and our experiments on DeepSeek-V3.2 and GLM-5.3 with an emulated scale-up memory pool show that \sys{} delivers up to 4.98$\times$ and 4.52$\times$ higher decode throughput than PCIe-based ECHO and HiSparse, respectively, and up to 1.34$\times$ over a naive integration on the same platform, across 16K--1M contexts.

%% file: sections/introduction.tex
\section{Introduction}
\label{sec:introduction}

Sparse attention is increasingly adopted in production LLMs, including recent DeepSeek~\citep{deepseekv32} and GLM~\citep{glm5} models, to improve long-context inference efficiency. These models use an indexer key (K-index) cache of compact keys to select historical tokens during decoding. By attending to only the selected tokens, they reduce attention computation and KV-cache accesses. Serving systems still need to retain the full KV history to keep all historical tokens available for attention. This history grows with context length, making GPU high-bandwidth memory (HBM) capacity a bottleneck for serving concurrency.

To alleviate HBM pressure, existing offloading systems such as HiSparse~\citep{hisparse} and ECHO~\citep{echo} offload the full KV cache to host memory while retaining the K-index cache and hot KV buffers for frequently accessed entries in GPU HBM.
During decoding, the system fetches missing KV entries from host memory on demand before executing attention. However, in PCIe-based architectures, limited interconnect bandwidth inflates decode latency, and host memory capacity restricts request concurrency, ultimately constraining decode throughput.

Recent innovations in scale-up interconnects enable scale-up memory pools, such as X-Mem~\citep{netpreme_xm100,netpreme2026xmem,netpreme2026xmsparse}, to address both limits. These pools offer both higher bandwidth than PCIe-attached CXL~\citep{cxl2020} and larger capacity than NVLink-C2C~\citep{nvlinkc2c2022} (Table~\ref{tab:memory-expansion}). We take one step further and ask: \emph{How can we maximize serving throughput of sparse-attention LLMs with scale-up memory pools?}

To understand the design space, we conduct an in-depth analysis to characterize sparse attention atop scale-up memory pools (\secref{sec:analysis}).
We find that decoding throughput is bounded by HBM capacity even after KV-cache offloading, as hot KV buffers and a fully resident K-index cache occupy substantial HBM. Moreover, KV-cache and K-index accesses exhibit distinct patterns, calling for tailored I/O mechanisms (e.g., load/store, TMA, and DMA) for efficient offloading.

Based on these insights, we present \sys{}, the first sparse-attention offloading system over scale-up memory for high-throughput LLM serving (\secref{sec:design}). \sys{} aggressively offloads the full KV and K-index caches to external memory, keeping only compact hot KV buffers, K-index staging buffers, and a subset of K-index pages in HBM.
In effect, \sys{} trades scale-up fabric bandwidth for HBM capacity, enabling larger batches and higher decode throughput.

To make \sys{} practical, we face two critical design challenges: (1) how to retrieve missing KV entries without exposing additional I/O latency, and (2) how to enable high-bandwidth K-index cache prefetching without interfering with decoding.

For the first challenge, we design a fused retrieval-attention kernel that overlaps on-demand KV fetch with attention computation at tile-level granularity (\secref{sec:design-fused-attention}). \sys{} exploits the independence of KV tiles in Split-K attention to schedule resident tiles first. A bounded tile pipeline overlaps attention on one tile with TMA retrieval of subsequent tiles as shared-memory buffers become available. Partial attention results are merged to produce the final output. By eliminating the separate swap-in stage before attention, \sys{} reduces exposed retrieval latency and supports smaller hot KV buffers for larger batches.

For the second challenge, we design SM-efficient GPU-initiated DMA for K-index prefetching and priority scheduling to mitigate traffic contention on the scale-up fabric (\secref{sec:design-index-prefetch}). We find that CPU submission overhead limits effective DMA bandwidth for small K-index page transfers. \sys{} therefore generates transfer descriptors on the GPU using only one SM and batches doorbell updates on the CPU, while leveraging GPU copy engines to handle data movement. Furthermore, to mitigate bandwidth contention on the shared scale-up fabric, \sys{} prioritizes on-demand KV retrieval and model collectives over K-index prefetching by pausing the copy engine at semaphore checkpoints in the DMA command stream.

We have implemented \sys{} on top of SGLang~\citep{sglang} and evaluate it on DeepSeek-V3.2 and GLM-5.3 using the X-Mem VM~\citep{xmemvm} across context lengths from 16K to 1M (\secref{sec:evaluation}). Our experimental results show that \sys{} achieves up to 4.98$\times$ and 4.52$\times$ the decode throughput of PCIe-based ECHO and HiSparse, respectively, and up to 1.34$\times$ that of a naive integration of HiSparse on the same platform.

Our contributions are:
\vspace{-1em}
\begin{itemize}
\setlength{\itemsep}{0pt}
    \item We characterize sparse-attention offloading over scale-up memory pools and show that HBM capacity, occupied by hot KV buffers and the resident K-index cache, remains the bottleneck after KV-cache offloading.
    \item We present \sys{}, the first sparse-attention offloading system over scale-up memory, which offloads both the KV and K-index caches. We design a fused retrieval-attention kernel for KV offloading, along with GPU-initiated DMA and priority scheduling for K-index offloading.
    \item We implement \sys{} atop SGLang and demonstrate its throughput gains on DeepSeek-V3.2 and GLM-5.3 through comparative evaluations and ablation studies.
\end{itemize}
\vspace{-1em}

%% file: sections/background.tex
\section{Background}
\label{sec:background}

\subsection{Sparse Attention and KV-Cache Offloading}
\label{sec:background:offloading}

Dynamic sparse attention selects a query-dependent subset of historical tokens at each decoding step~\citep{quest,twilight}.
Selection uses an indexer key (K-index) cache of compact key representations or metadata to score historical tokens or pages.
For example, DeepSeek-V3.2 and GLM-5 use lightweight indexers with compact per-token keys to select the $k$ highest-scoring tokens~\citep{deepseekv32,glm5}.
At each decoding step, selection reads the full K-index cache, whereas attention accesses only the selected KV entries.
Sparse selection reduces per-step attention computation and KV accesses, but the full KV history is retained because tokens excluded from the current selection may be needed in later steps.
The KV-cache storage requirement therefore grows with context length, limiting serving concurrency when the full history is kept in GPU HBM.

To alleviate this HBM capacity bottleneck, systems such as SparseServe, HiSparse, and ECHO offload the full KV cache to host memory while retaining the full K-index cache and hot KV buffers in GPU HBM~\citep{sparseserve,hisparse,echo}.
This reduces the per-request HBM footprint and supports larger decode batches.
Consecutive decoding steps often select many of the same tokens, so caching recently accessed KV entries in these buffers reduces host-to-device transfers~\citep{ess}.
Selected entries missing from the buffers are fetched on demand in small blocks from scattered host-memory addresses before attention begins.
GPU kernels perform these fine-grained transfers over PCIe using load/store instructions executed on a small number of streaming multiprocessors (SMs), which suffice to saturate PCIe bandwidth.

\subsection{Decode Parallelism}
\label{sec:background:decode-parallelism}

We consider decoding with data-parallel attention and expert-parallel mixture-of-experts (MoE) layers.
With attention data parallelism (DP), each GPU processes a different subset of requests.
With expert parallelism (EP), experts are distributed across GPUs, requiring communication to dispatch token activations to the selected experts and combine their outputs.

\subsection{GPU Memory Expansion Technologies}
\label{sec:background:memory-expansion}

We review hardware choices for offloading and summarize their bandwidth/capacity trade-offs in Table~\ref{tab:memory-expansion}.

\input{sections/table_code/memory_expansion}

PCIe-based systems use host memory for offloading~\citep{hisparse}.
For example, an eight-GPU DGX B300 server has 2~TB of host memory by default~\citep{dgxb300userguide}.
Compute Express Link (CXL) expands host memory to tens of terabytes through memory expanders and pooling~\citep{marvell_cxl}.
In both cases, transfers to PCIe-attached GPUs are limited by the GPU's PCIe bandwidth~\citep{pcisig_pcie5,lmb}.

NVIDIA superchips couple CPUs and GPUs through the high-bandwidth, cache-coherent NVLink Chip-to-Chip (C2C) interconnect, enabling direct GPU access to CPU-attached LPDDR memory at higher bandwidth than PCIe~\citep{superoffload,nvlinkc2c2022}.
Offloading capacity is bounded by CPU-attached memory; for example, the GB200 Grace Blackwell Superchip provides up to 480~GB of LPDDR5X memory shared by two GPUs~\citep{nvidia_gb200}.

Scale-up memory pools connect dedicated memory nodes to GPUs through the scale-up fabric (e.g., NVLink).
The pooled memory is exposed in the GPU address space, supporting peer-to-peer access without host-memory staging.
Each memory node can provide tens of terabytes of DDR5 memory, and GPUs access it at scale-up fabric bandwidth.
For example, each X-Mem~\cite{netpreme_xm100} node is designed to provide up to 24~TB of DDR5 memory and approximately 4~TB/s of aggregate memory bandwidth.
Adding nodes scales both capacity and aggregate bandwidth independently of GPU count. Any GPU in the scale-up domain can access any memory node over the NVSwitch fabric.
In scale-up memory pools, offloading traffic shares the fabric with model collectives.

\parab{GPU I/O primitives.}
GPUs access these memory tiers through three I/O mechanisms.
With GPU load/store, SM threads load remote data into registers and store it in local HBM~\citep{strata}.
SM threads can also issue asynchronous transfers between global and shared memory through the Tensor Memory Accelerator (TMA).
A single thread can initiate a bulk tensor copy, with hardware handling tensor address generation and data movement.
This reduces instruction overhead and avoids staging the payload in registers~\citep{nvidia_hopper_tuning}.
For direct memory access (DMA), the CPU driver enqueues transfer commands and rings a doorbell to notify the GPU of pending work~\citep{yan2026commandstreams}.
Dedicated copy engines then move the data without an SM copy kernel~\citep{nvidia_cuda_programming}.

%% file: sections/table_code/memory_expansion.tex
\begin{table}[t]
    \centering
    \caption{Trade-off between bandwidth and capacity.}
    \label{tab:memory-expansion}
    \small
    \setlength{\tabcolsep}{4pt}
    \renewcommand{\arraystretch}{1.1}
    \begin{tabular*}{\columnwidth}{@{\extracolsep{\fill}}lrl@{}}
        \toprule
        \textbf{Approach} & \textbf{BW/GPU} & \textbf{Capacity (scope)} \\
        \midrule
        PCIe 5.0 $\times$16          & 64~GB/s  & 2~TB (8-GPU server) \\
        CXL 2.0 $\times$16           & 64~GB/s  & Tens of TB (pool) \\
        NVLink-C2C (GB200)           & 225~GB/s & 480~GB (2 GPUs) \\
        Scale-up pool (e.g., X-Mem)  & 900~GB/s & Tens of TB (node) \\
        \bottomrule
    \end{tabular*}
\end{table}

%% file: sections/analysis.tex
\section{In-Depth Analysis}
\label{sec:analysis}

\input{sections/fig_code/analysis/bandwidth_capacity}

We examine the bandwidth and capacity bottlenecks of sparse-attention decoding, characterize the memory footprint and access patterns of its caches, and evaluate I/O mechanisms for offloading.

Unless otherwise stated, we profile HiSparse with the full GLM-5.3-FP8 model on four NVIDIA B300 GPUs at 128K context~\citep{glm5}.
We use the publicly available X-Mem VM~\citep{xmemvm}, which emulates a scale-up memory pool with four additional B300 GPUs serving as memory nodes over NVLink (denoted Scale-up Pool).
PCIe-Host and PCIe-CXL retrieve KV data from host memory at PCIe Gen5 bandwidth.
We emulate C2C by pacing NVLink KV reads at 225~GB/s per GPU.
We emulate capacities beyond the 1~TB of physical backing by repeatedly mapping physical host memory for CXL or GPU memory for Scale-up Pool into a larger virtual address space.
Each request uses a 4K-entry hot KV buffer per layer and selects 2K KV entries per layer per step.
Unless otherwise specified, we use a 15\% miss rate, a typical ratio consistent with our observations as well as prior work~\citep{hisparse}.

\subsection{Bottleneck Analysis and Hardware Opportunity}
\label{sec:analysis-bottlenecks}

\input{sections/fig_code/analysis/memory_footprint}
\input{sections/fig_code/analysis/hot_buffer_tradeoff}

Decode throughput is constrained by both interconnect bandwidth and memory capacity.
Bandwidth affects the cost of retrieving missing KV entries, while HBM and external-memory capacities jointly limit the batch size.
We examine these constraints through testbed measurements across memory-expansion configurations.
Each configuration uses a 60~GiB HBM cache budget per GPU, excluding model weights and runtime allocations.

\parab{Bottleneck analysis.}
Figure~\ref{fig:analysis-bandwidth-capacity} compares configurations to examine how external-memory capacity and interconnect bandwidth constrain decode throughput.
At 16K, 1~TB of external memory already supports the HBM-limited batch of 241 per GPU (Figure~\ref{fig:analysis-bandwidth-capacity-16k}).
PCIe-Host and PCIe-CXL overlap because they share the same PCIe bandwidth and HBM-limited maximum batch size.
At this batch, Scale-up Pool provides 1.92$\times$ the decode throughput of PCIe-Host and PCIe-CXL; additional external capacity does not increase the feasible batch size.

At 128K, the 1~TB pools in the PCIe-Host and C2C configurations cap the batch at 36 per GPU, so C2C's higher bandwidth cannot enable larger batches (Figure~\ref{fig:analysis-bandwidth-capacity-128k}).
Expanding capacity to 24~TB with CXL permits a batch of 109 per GPU, a 3.03$\times$ increase in batch size while retaining the PCIe Gen5 host-memory access path.
Longer contexts amplify this capacity benefit: at 1M, the same expansion enables 5$\times$ larger batches and 2.99$\times$ higher throughput (Appendix~A.1).

\parab{Hardware opportunity.}
Scale-up memory pools alleviate both constraints through their large capacity and high-bandwidth interconnect.
At 128K and a batch size of 109 per GPU, Scale-up Pool achieves 1.71$\times$ the decode throughput of CXL.
Its throughput reaches approximately 94.2\% of the compute-only counterpart.
However, GPU-resident caches exhaust the HBM budget, limiting further increases in batch size even though only about 3~TB of the 24~TB external memory pool is used.

\begin{samepage}
\noindent\textbf{Takeaway.} Scale-up memory pools shift the capacity bottleneck to HBM. Further throughput gains require freeing HBM for larger batches via additional offloading.
\par
\end{samepage}

\subsection{Memory Characteristics}
\label{sec:analysis-memory}

\input{sections/fig_code/analysis/swap_in_volume_latency}
\input{sections/fig_code/analysis/io_mechanisms}

To identify offloading opportunities in sparse-attention models, we investigate the hot KV buffers and the K-index cache.
As shown in Figure~\ref{fig:analysis-memory-characteristics}, we observe that even after KV cache offloading, this resident state consumes approximately 40\% of GPU memory for resident states alongside the model weights.
We further analyze their memory footprints to assess potential HBM savings, and evaluate their access patterns to determine the I/O requirements for offloading.

\parab{Hot KV buffers.} Existing sparse-attention offloading systems typically configure hot KV buffers at $2$--$4\times$ the Top-K token budget (4K--8K entries for a 2K-token selection) to retain frequently accessed tokens across steps and reduce retrieval volumes for missed entries~\citep{sparseserve,hisparse,sac}.
We find that such configurations are largely over-provisioned in the high-bandwidth regime of scale-up memory pools. As shown in Figure~\ref{fig:analysis-hot-buffer-miss-rate}, reducing these buffers to 2K saves 50--75\% of their HBM footprint per request. Although this reduction increases the miss rate from approximately 4--10\% to 25\%, the additional misses have only a modest performance impact on Scale-up Pool. Specifically, at a 128K context length and a batch size of 64 per GPU, increasing the miss rate from 5\% to 30\% only raises the decode-step latency by 10.0\% and reduces throughput by 9.1\% (Figure~\ref{fig:analysis-miss-rate-throughput}).

Missing KV entries are dynamically determined by layer-wise token selection and must be retrieved before attention computation.
We observe that on-demand KV retrieval in HiSparse\footnote{\href{https://github.com/sgl-project/sglang/blob/5469faec45c89110be1ebc3828ee409932077435/python/sglang/kernels/jit/csrc/hisparse.cuh\#L367-L687}{HiSparse swap-in kernel on GitHub}} is bounded by latency, not the interconnect bandwidth. Our results reveal that a $32\times$ increase in batch size (from 1 to 32 per GPU) increases the KV payload proportionally, but only raises the mean swap-in latency by 6.5\% (74.1 to 78.9~$\mu$s; Figure~\ref{fig:analysis-swap-in-latency}).
This finding also echoes our earlier observation in Figure~\ref{fig:analysis-miss-rate-throughput}.
The root cause is twofold. First, the swap-in kernel launches only a single thread block per request and fetches entries sequentially via warps. This severely restricts SM parallelism and imposes significant retrieval latency on the decode critical path.
In addition, retrieved KV entries are staged in HBM before being loaded into shared memory for computation, introducing extra data movement overhead on the critical path.
Overcoming these limitations requires eliminating redundant data copies and effectively overlapping KV retrieval with computation.
Thus, KV retrieval favors I/O mechanisms that offer low latency and support overlap with computation.

\parab{K-index cache.}
The K-index cache retains index entries for the full token history of every request, causing its HBM footprint to scale with both context length and batch size.
These entries are stored in pages, each holding index keys for a fixed number of tokens.
Currently, systems like HiSparse keep all these index pages permanently resident in HBM, even though each layer only requires its corresponding pages during token selection.
Consequently, at a 128K context length, this cache occupies approximately 33.7\% and 24.9\% of total HBM in the illustrative DeepSeek-V3.2 and GLM-5.3 configurations, respectively (Figure~\ref{fig:analysis-memory-characteristics}).
Offloading index pages can therefore free up valuable HBM capacity to support larger batch sizes.

Unlike dynamical KV cache accesses, token selection requires a deterministic, full-history scan of the K-index at every decode step.
This predictability allows us to offload these pages and prefetch them in advance.
K-index offloading is highly bandwidth-sensitive; because prefetched pages must arrive during the preceding computation, the offloadable volume is strictly bounded by available bandwidth.
For a single request, fully offloading the K-index in the GLM-5.3 configuration (128K context) transfers approximately 363~MB per step, compared with 15.7~MB of KV-miss payload at a 15\% miss rate.
Therefore, it favors I/O mechanisms with high bandwidth and low interference with concurrent compute kernels.

\noindent\textbf{Takeaway.}
Shrinking hot KV buffers and offloading the K-index cache can free HBM for larger batches.
Realizing this opportunity requires low-interference, overlap-friendly I/O mechanisms, with low latency for KV retrieval and high bandwidth for K-index prefetching.

\subsection{GPU I/O Mechanisms}
\label{sec:analysis-io}

To identify suitable I/O mechanisms for KV-cache and K-index offloading, we compare GPU-native data movement mechanisms, including load/store, TMA, and DMA, in terms of I/O performance and interference with concurrent decoding.
The background decoding workload uses 128K context length, and a batch size of 64 per GPU.
The message granularity is typically 656~B or 1~KB for KV-cache accesses and 8~KB or 16~KB for K-index accesses.

\noindent\textbf{GPU load/store.}
Increasing the number of copy blocks raises memory-level parallelism and transfer bandwidth, but also increases competition with decoding for SM resources.
For 656~B KV transfers, 96 copy blocks achieve 483~GB/s but reduce concurrent decode throughput by 45.2\% (Figure~\ref{fig:analysis-sm-contention}).
We observe that the mean per-message latency remains nearly constant at approximately 3.8~$\mu$s as the number of copy blocks increases.
While providing low latency, load/store causes substantial interference with concurrent computation, making it less suitable for offloading I/O.

\noindent\textbf{TMA.}
TMA fetches KV tiles directly into shared memory, allowing attention to consume them within the same kernel without intermediate HBM staging.
TMA provides bandwidth comparable to load/store while causing less interference with decoding.
For the same 656~B transfers, TMA achieves 394~GB/s with 16 copy blocks while reducing decode throughput by 6.3\% (Figure~\ref{fig:analysis-tma-contention}).
Increasing TMA parallelism to 32 copy blocks raises bandwidth to 612~GB/s, but the throughput reduction also grows to 29.4\%.
We observe that the mean latency remains stable at 1.9~$\mu$s for a small number of copy blocks ($<16$), but increases to 4.3~$\mu$s when the block count exceeds 32.
TMA still competes with decoding for SM resources through its issuing warps and shared-memory buffers.
Overall, TMA is a good fit for low-latency KV retrieval, but its SM requirements make it less attractive for high-bandwidth K-index prefetching.

\noindent\textbf{DMA.}
DMA bandwidth for small transfers is limited by CPU submission and copy-engine execution.
For representative 16-KiB K-index pages, continuous CPU submission achieves only 22.9~GB/s, compared with 632~GB/s for prequeued execution without concurrent decoding (Figure~\ref{fig:analysis-dma-submission}).
Descriptor submission and doorbell writes limit the submission rate to approximately 1.4~million transfers per second.
For 656~B transfers, prequeued execution increases bandwidth from 0.92 to 58.8~GB/s, but remains far below the roughly 830~GB/s bulk-transfer bandwidth, indicating an additional limit in copy-engine execution.
For 656-B transfers with concurrent decoding, the mean latency from CPU submission to completion is 26.2~$\mu$s, compared with 8.5~$\mu$s for prequeued execution.
These submission overheads and high startup latencies make DMA unsuitable for fine-grained, on-demand KV retrieval.

In terms of contention, although DMA operations do not consume SM resources, bulk transfers still interfere with decoding.
With 1-GiB transfers sustaining approximately 808~GB/s, decode throughput decreases by 11.8\%.
A separate kernel-level profile reveals that this interference primarily stems from traffic contention: NCCL kernel time for MoE dispatch and combine increases by 46.2\%, compared with 4.1\% for compute kernels (Figure~\ref{fig:analysis-dma-degradation}).
Despite its inefficiency with small transfers, DMA remains appealing for K-index prefetching due to its minimal SM interference.
Efficient submission and coordination with model communication are therefore needed.

\begin{samepage}
\noindent\textbf{Takeaway.}
TMA is suitable for on-demand KV retrieval. DMA is appealing for K-index prefetching, but requires efficient submission and coordination with model communication.
\par
\end{samepage}

%% file: sections/fig_code/analysis/bandwidth_capacity.tex
\begin{figure}[t]
    \centering
    \includegraphics[width=\linewidth]{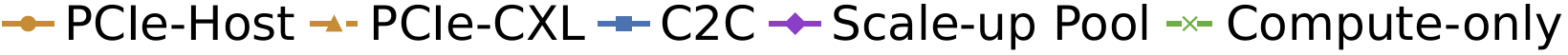}\par
    \vspace{2pt}
    \begin{subfigure}[t]{0.49\linewidth}
        \centering
        \includegraphics[width=\linewidth]{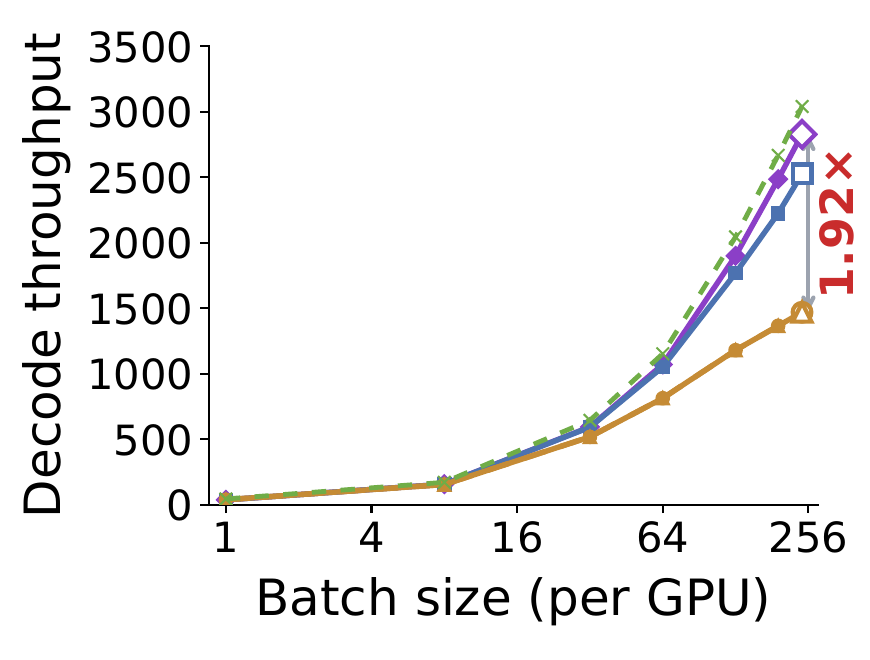}
        \caption{16K context.}
        \label{fig:analysis-bandwidth-capacity-16k}
    \end{subfigure}\hfill
    \begin{subfigure}[t]{0.49\linewidth}
        \centering
        \includegraphics[width=\linewidth]{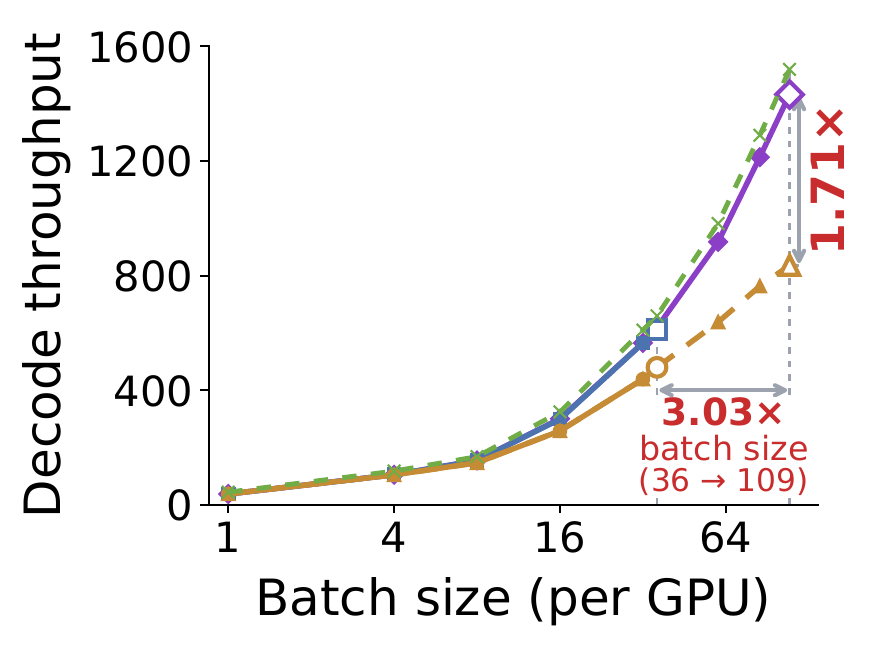}
        \caption{128K context.}
        \label{fig:analysis-bandwidth-capacity-128k}
    \end{subfigure}
    \caption{Decode throughput across memory-expansion configurations. Throughput is in tokens/s/GPU; external capacity is shared by four GPUs. ``Compute-only'' disables KV transfers.}
    \label{fig:analysis-bandwidth-capacity}
\end{figure}

%% file: sections/fig_code/analysis/memory_footprint.tex
\begin{figure}[t]
    \centering
    \includegraphics[width=\linewidth]{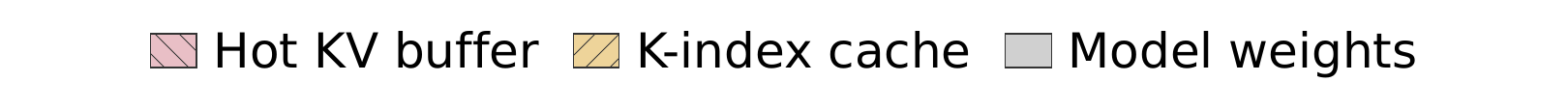}\par
    \vspace{2pt}
    \begin{subfigure}[t]{0.49\linewidth}
        \centering
        \includegraphics[width=\linewidth]{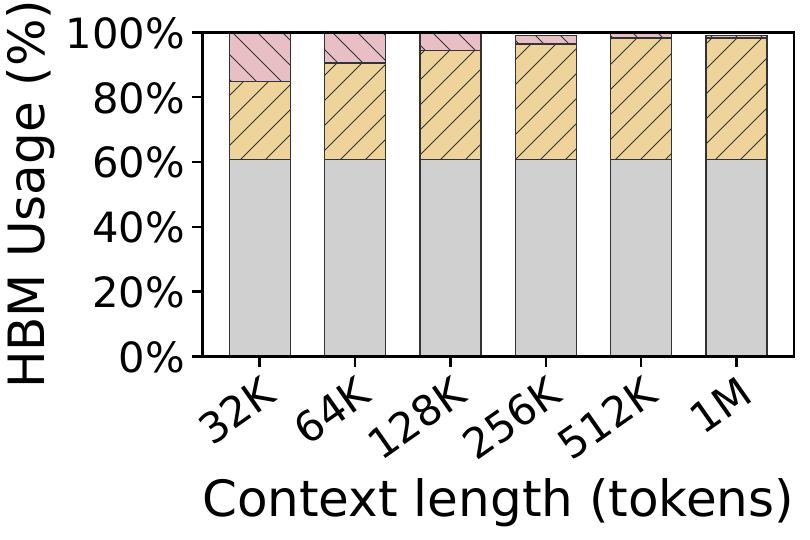}
        \caption{DeepSeek-V3.2.}
        \label{fig:analysis-hbm-deepseek}
    \end{subfigure}\hfill
    \begin{subfigure}[t]{0.49\linewidth}
        \centering
        \includegraphics[width=\linewidth]{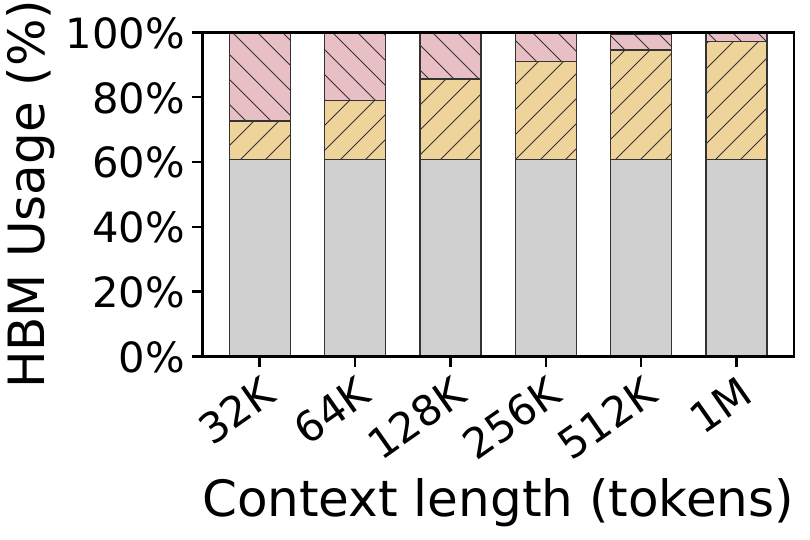}
        \caption{GLM-5.3.}
        \label{fig:analysis-hbm-glm}
    \end{subfigure}
    \caption{Illustrative HBM usage breakdown at the maximum batch for each context, assuming 288~GB per GPU and excluding runtime overheads.}
    \label{fig:analysis-memory-characteristics}

\end{figure}

%% file: sections/fig_code/analysis/hot_buffer_tradeoff.tex
\begin{figure}[t]
    \centering
    \begin{subfigure}[t]{0.49\linewidth}
        \centering
        \includegraphics[width=\linewidth]{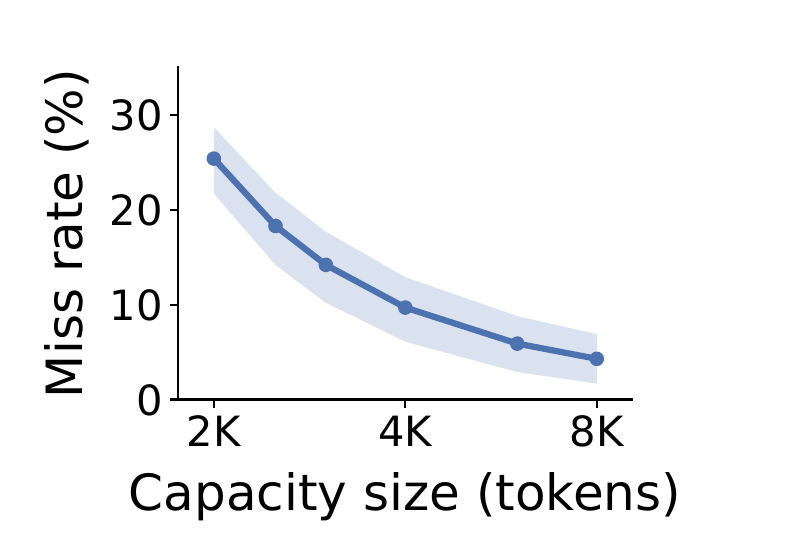}
        \caption{LongBench v2 miss rate.}
        \label{fig:analysis-hot-buffer-miss-rate}
    \end{subfigure}\hfill
    \begin{subfigure}[t]{0.49\linewidth}
        \centering
        \includegraphics[width=\linewidth]{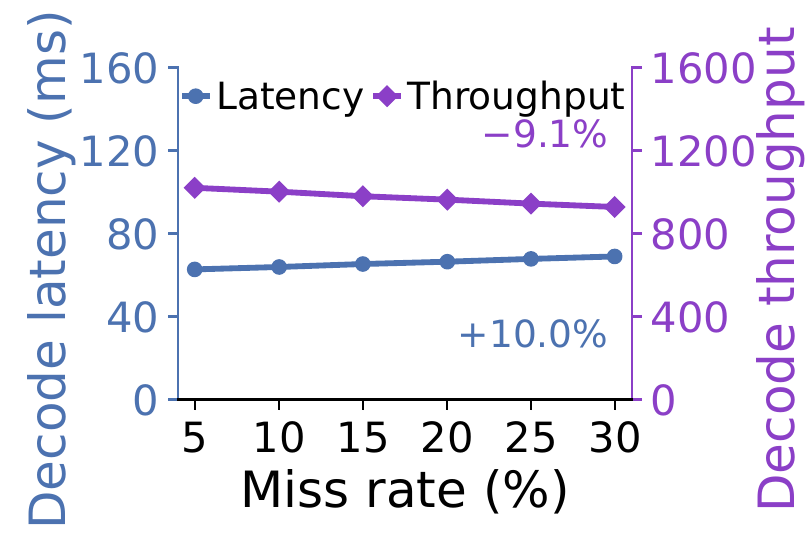}
        \caption{Miss-rate sensitivity.}
        \label{fig:analysis-miss-rate-throughput}
    \end{subfigure}
    \caption{Hot-buffer tradeoffs. (a) Digitized LongBench v2 miss rates; shading spans five input-length groups. (b) Measured Scale-up Pool decode latency and throughput (tokens/s/GPU) at 128K context and batch 64 per GPU, showing 5\%--30\% misses. Annotations compare 30\% with 5\% misses.}
    \label{fig:analysis-hot-buffer-tradeoff}

\end{figure}

%% file: sections/fig_code/analysis/swap_in_volume_latency.tex
\begin{figure}[t]
    \centering
    \begin{subfigure}[t]{0.49\linewidth}
        \centering
        \includegraphics[width=\linewidth]{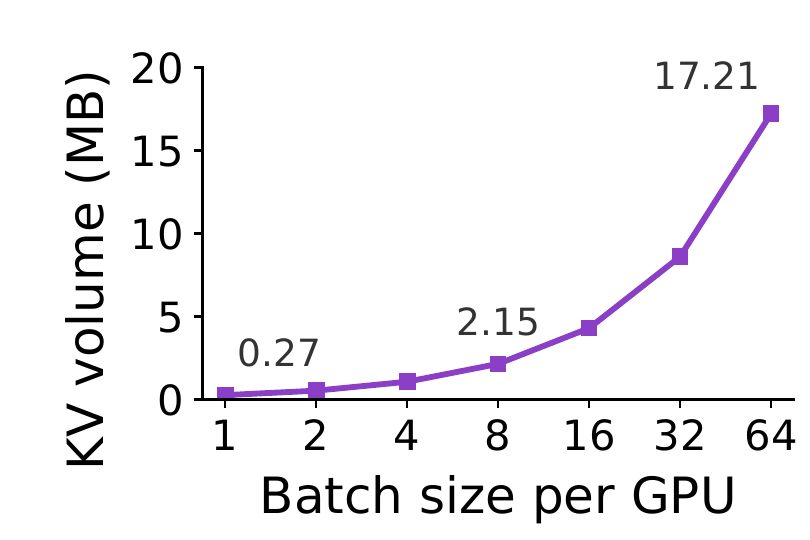}
        \caption{Requested KV volume.}
        \label{fig:analysis-batch-kv-volume}
    \end{subfigure}\hfill
    \begin{subfigure}[t]{0.49\linewidth}
        \centering
        \includegraphics[width=\linewidth]{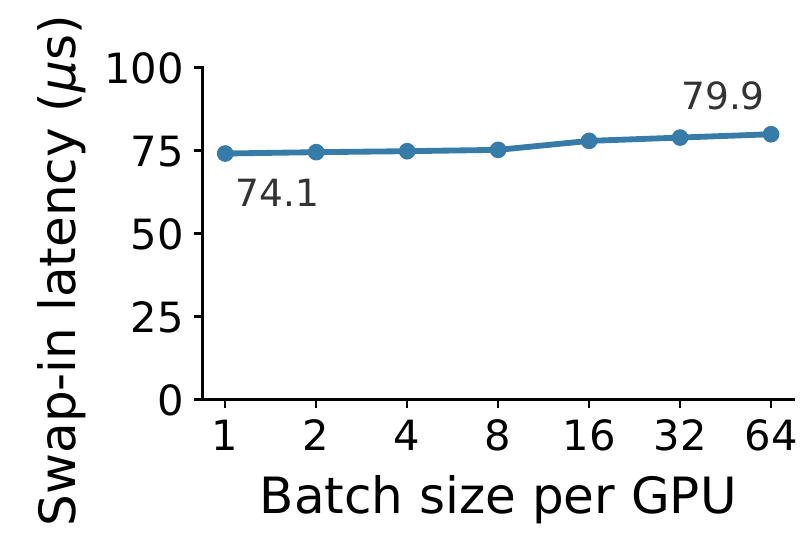}
        \caption{Swap-in kernel latency.}
        \label{fig:analysis-batch-kernel-latency}
    \end{subfigure}
    \caption{Scale-up Pool KV swap-in at 128K context and 20\% misses. Volume is per layer per GPU; latency is the mean kernel duration.}
    \label{fig:analysis-batch-swap-in-latency}
    \label{fig:analysis-swap-in-latency}
\end{figure}

%% file: sections/fig_code/analysis/io_mechanisms.tex
\begin{figure*}[t]
    \centering
    \makebox[0.4933\textwidth][c]{\includegraphics[height=24pt]{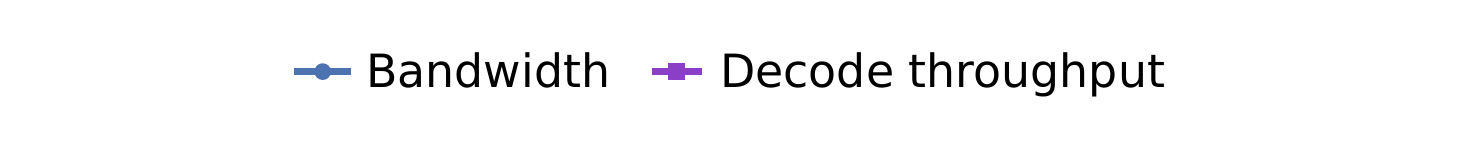}}\hfill
    \makebox[0.24\textwidth][c]{\includegraphics[height=24pt]{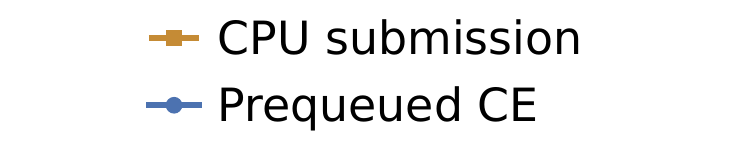}}\hfill
    \makebox[0.24\textwidth][c]{\includegraphics[height=24pt]{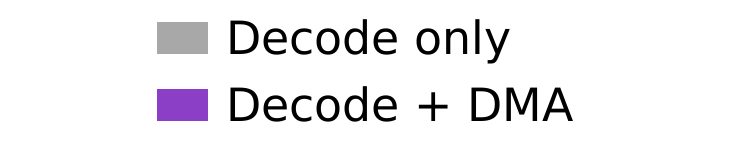}}\par
    \vspace{2pt}
    \begin{subfigure}[t]{0.24\textwidth}
        \centering
        \includegraphics[width=\linewidth]{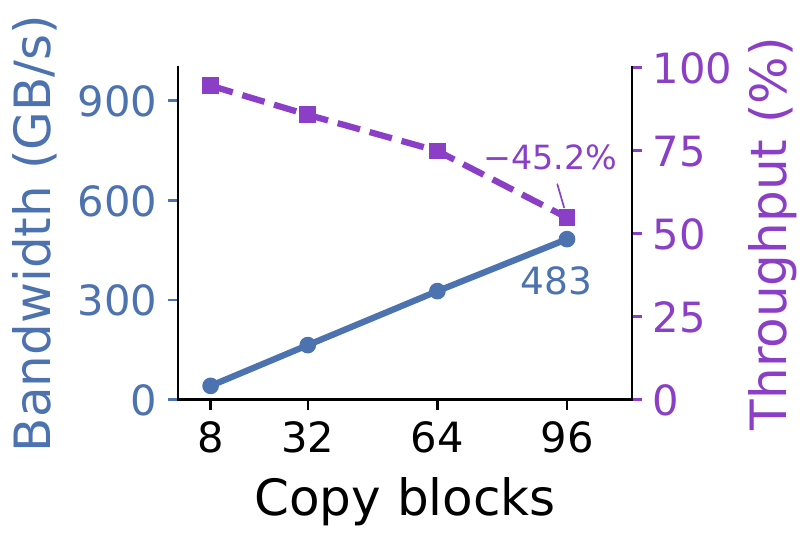}
        \caption{GPU load/store.}
        \label{fig:analysis-sm-contention}
    \end{subfigure}\hfill
    \begin{subfigure}[t]{0.24\textwidth}
        \centering
        \includegraphics[width=\linewidth]{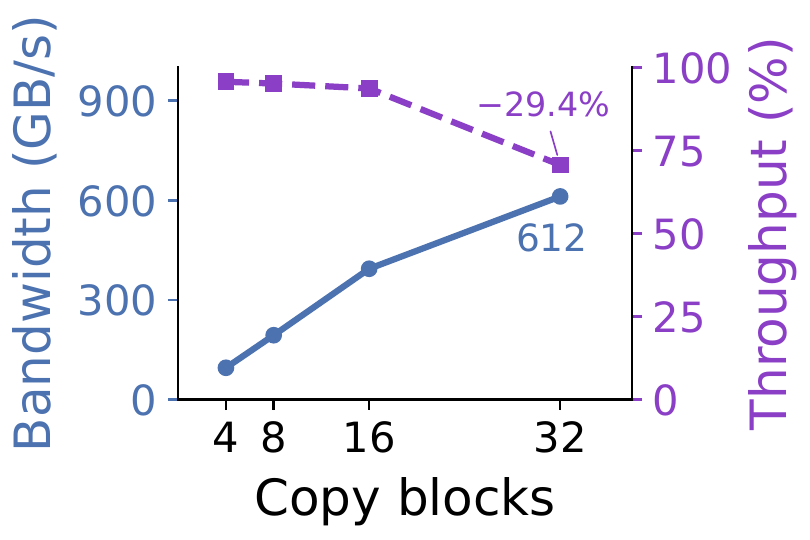}
        \caption{TMA.}
        \label{fig:analysis-tma-contention}
    \end{subfigure}\hfill
    \begin{subfigure}[t]{0.24\textwidth}
        \centering
        \includegraphics[width=\linewidth]{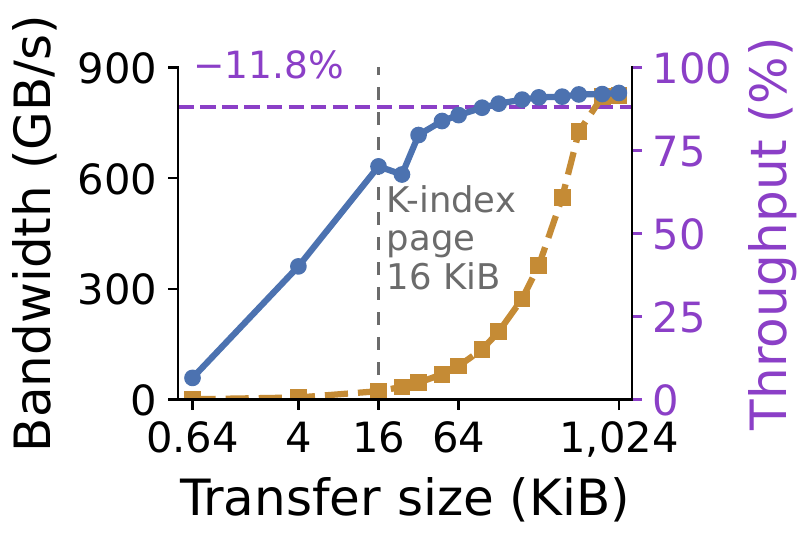}
        \caption{DMA.}
        \label{fig:analysis-dma-submission}
    \end{subfigure}\hfill
    \begin{subfigure}[t]{0.24\textwidth}
        \centering
        \includegraphics[width=\linewidth]{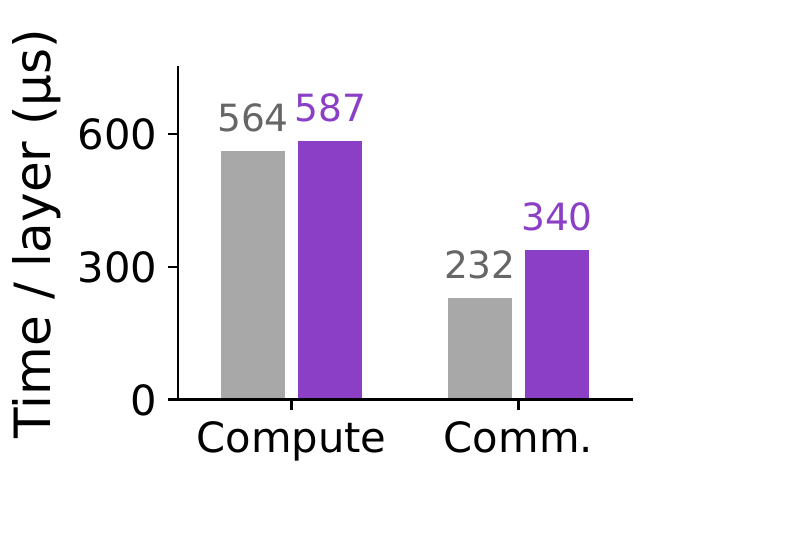}
        \caption{DMA interference.}
        \label{fig:analysis-dma-degradation}
    \end{subfigure}
    \caption{GPU I/O bandwidth and decode interference. Throughput is relative to matched no-transfer baselines. (a,b) 656-B reads with concurrent decoding at 128K context and batch 64 per GPU; purple annotations show throughput loss. (c) Prequeued copy-engine (CE) execution excludes CPU submission; the dashed line shows the separate 1-GiB bulk-DMA result (11.8\% throughput loss). (d) Per-layer kernel times with and without background 1-GiB DMA transfers at about 808~GB/s.}
    \label{fig:analysis-io-contention}
\end{figure*}

%% file: sections/design.tex
\section{Design and Implementation}
\label{sec:design}

\subsection{System Overview}
\label{sec:design-overview}

\input{sections/fig_code/design/system_overview}

\sys{} supports larger decode batches by reducing the HBM footprint of the KV and K-index caches.
It preserves the model's token-selection and sparse-attention semantics.
Following the access patterns identified in Section~\ref{sec:analysis-memory}, \sys{} uses fused retrieval and attention for on-demand KV accesses and prefetching for predictable K-index accesses.

The external memory pool provides backing storage for both caches, while HBM holds hot KV buffers, resident K-index pages, and reusable staging buffers.
Each request maintains a hot KV buffer per attention layer to retain recently selected entries, with its capacity fixed at startup.
New cache entries are first written to HBM and subsequently propagated to the memory pool.

The cache manager maintains data mappings and buffer allocations for both data paths.
It determines which K-index pages remain resident and which use staging buffers. The serving scheduler forms batches subject to the cache manager's HBM capacity and transfer budget checks.

Figure~\ref{fig:design-system-overview} summarizes the cache placement and the decode flow.
During each decode step, K-index prefetching advances alongside layer computation.
Once its required pages are ready, each indexer appends the current token's index key and selects tokens.
The selected token IDs then drive fused KV retrieval and attention.

\begin{itemize}[leftmargin=*]
    \setlength{\itemsep}{0pt}
    \item \textbf{Fused Retrieval and Attention} (Section~\ref{sec:design-fused-attention}).\nopagebreak[4]
    \sys{} uses smaller hot KV buffers to free HBM for larger decode batches.
    A fused kernel retrieves missing KV entries with TMA while computing attention over the selected tokens.
    \item \textbf{Efficient K-index Prefetching} (Section~\ref{sec:design-index-prefetch}).\nopagebreak[4]
    \sys{} prefetches offloaded K-index pages into reusable HBM staging buffers using \emph{GPU-initiated DMA}.
    A prefetch scheduler gives on-demand KV retrieval and model collectives precedence over K-index prefetching.
\end{itemize}

\subsection{Fused Retrieval and Attention}
\label{sec:design-fused-attention}

\input{sections/fig_code/design/fused_retrieval_timeline}

Smaller hot KV buffers increase on-demand misses, exposing more retrieval latency in a separate swap-in stage before attention (Section~\ref{sec:analysis-memory}).
Fusion alone does not remove this data dependency; \sys{} reorders KV tiles and pipelines retrieval with attention computation.

\parab{Resident-first traversal.}
The kernel uses online softmax~\citep{onlinesoftmax,flashattention} to accumulate attention over the selected KV tiles.
It traverses resident tiles first so that retrieval of missing tiles can overlap attention computation.

After token selection, the cache manager constructs separate lists of hot-buffer hits and misses.
Hit entries reference HBM locations, while miss entries reference memory-pool locations and include destination slots for cache refill.
The kernel groups these entries into tiles and places resident tiles before missing tiles in the logical traversal order.
Split-K partitions this ordered sequence into contiguous portions for parallel processing by thread blocks.
Within each block, resident tiles precede missing tiles.
A miss-only block waits for its first tile, then overlaps attention on the current tile with retrieval of subsequent tiles.
Partial outputs are merged after the splits finish, using their softmax normalization statistics.

\parab{Tile-level pipeline.}
To overlap retrieval with computation, the kernel assigns retrieval, dequantization, and attention to separate warps and warpgroups connected through double-buffered shared memory.
Producer warps issue TMA gathers from the appropriate HBM or memory-pool source into shared memory.
Remote KV tiles are consumed directly from shared memory, while HBM refill proceeds alongside dequantization.
While attention consumes one tile, the retrieval and dequantization stages prepare subsequent tiles as buffer space permits.

Stages consume only ready tiles.
Shared-memory buffers are reused only after computation and refill finish reading them.
The available buffers bound how far retrieval can run ahead; a late tile stalls its consumer.
Figure~\ref{fig:design-fused-retrieval-timeline}(a) contrasts this pipeline with a separate swap-in stage.

\parab{Cache refill.}
The kernel refills the hot KV buffer from the retrieved tiles without additional memory-pool reads (Figure~\ref{fig:design-fused-retrieval-timeline}(b)).
The buffer is sized to accommodate a full Top-K selection, allowing all misses to be refilled while retaining the current hits.
The cache manager follows HiSparse's LRU policy, selecting refill slots from entries not used by the current attention invocation.
It updates the cache mappings and LRU state when constructing the KV access plan.
The kernel writes raw quantized bytes to the designated refill slots during dequantization and preserves the original scales.
Subsequent accesses to the updated entries are ordered after the fused kernel completes.

\subsection{Efficient K-index Prefetching}
\label{sec:design-index-prefetch}

\input{sections/fig_code/design/k_index_prefetch}

Offloading K-index pages reduces HBM usage but introduces transfers that must complete before token selection.
The predictable access order enables prefetching during preceding computation (Section~\ref{sec:analysis-memory}).
\sys{} combines on-demand allocation and offloading, GPU-initiated DMA, and priority scheduling to exploit this opportunity.

\subsubsection{On-Demand Allocation and Offloading}
\label{sec:design-prefetch-staging}

The cache manager allocates HBM for K-index pages on demand and keeps them resident within its HBM budget.
When additional allocations would exceed this budget, it offloads enough complete pages to accommodate new requests or growing contexts.
The manager distributes the required offloading across requests in proportion to their complete page counts.
Within each request, it selects later complete pages first and uses the same page selection across indexer layers.
These pages are prefetched into reusable staging buffers before token selection at each decode step.
As space becomes available within this budget, the manager gradually restores offloaded pages to resident storage.
Each request's partially filled page remains resident.
A full page becomes eligible for offloading only after it has been written to the memory pool.

Before a request joins the batch, the cache manager checks that the resulting allocation and transfers fit the HBM and transfer budgets.
The transfer check covers both recurring prefetches and initial page loads.
The transfer budget is the measured duration of the previous decode step.
The manager estimates transfer time from the transfer volume and measured copy bandwidth.
Requests exceeding either budget remain queued.

The manager alternates between two sets of HBM staging buffers across indexer invocations.
Each set can be reused after its previous indexer finishes reading.
Prefetches follow indexer execution order across decode steps and overlap computation outside fused attention and the protected collectives (Section~\ref{sec:design-communication-coordination}).
Because complete pages do not change across steps, early-layer prefetches for continuing requests overlap computation at the end of the preceding step.
The indexer accesses resident and staged pages through a common page table and reads only after the required transfers complete.
Figure~\ref{fig:design-index-prefetch}(a) illustrates the cross-layer schedule and buffer reuse.

\subsubsection{GPU-Initiated DMA}
\label{sec:design-gpu-dma}

Page-granular offloading produces many small transfers.
As Section~\ref{sec:analysis-io} shows, CPU submission can limit throughput well below the copy engine's execution bandwidth.
\sys{} moves per-transfer command generation to the GPU while retaining batch-level CPU publication.

A GPU producer running as one thread block on a single SM reads the cache manager's copy lists and generates DMA commands mapping memory-pool source pages to HBM destinations.
After making the commands visible, the producer updates a readiness counter in pinned host memory.
A CPU worker polls this counter, publishes ready command batches to the copy-engine queue, and rings the doorbell.
The copy engine transfers the payload and signals completion.
CPU publication work therefore scales with command batches rather than individual pages, while data movement uses copy engines.
The transfer engine separates initiation from completion, allowing the manager to launch prefetches ahead of layer execution and attach completion dependencies to the consuming indexers.
Figure~\ref{fig:design-index-prefetch}(b) shows this division of work.

\subsubsection{Priority Scheduling}
\label{sec:design-communication-coordination}

K-index prefetches share the scale-up fabric with on-demand KV retrieval and model collectives (Section~\ref{sec:analysis-io}).
The prefetch scheduler gives these two types of traffic priority over K-index prefetching.

It pauses prefetching during fused attention, MoE dispatch, and MoE combine, and resumes it after each operation completes.
GPU kernels on the model's CUDA stream update a semaphore in GPU memory before and after each protected operation.
The producer inserts semaphore-acquire commands at the start of each command batch and at regular intervals in the DMA command stream.
The copy engine pauses at the next checkpoint; any transfer segment already in progress completes normally.
Checkpoints are spaced so that each segment's transfer time is short relative to the protected operations.
The copy engine checks the semaphore directly, so pausing and resuming require no CPU round trip.

\subsection{Implementation}
\label{sec:design-implementation}

We have implemented \sys{} with approximately 8K lines of Python and C++/CUDA code on top of SGLang~\citep{sglang}, including a modified FlashMLA kernel and a GPU-initiated DMA library.

%% file: sections/fig_code/design/system_overview.tex
\begin{figure}[t]
    \centering
    \includegraphics[width=\linewidth]{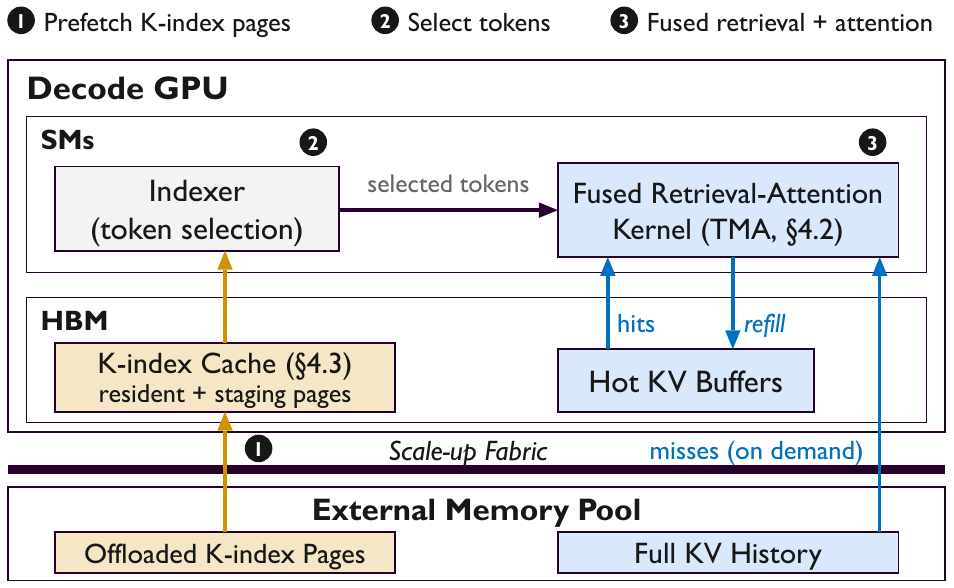}
    \caption{System overview of \sys{}. The fused kernel retrieves missing KV entries on demand, while DMA prefetches K-index pages into reusable HBM staging buffers.}
    \label{fig:design-system-overview}
\end{figure}

%% file: sections/fig_code/design/fused_retrieval_timeline.tex
\begin{figure}[t]
    \centering
    \includegraphics[width=\linewidth]{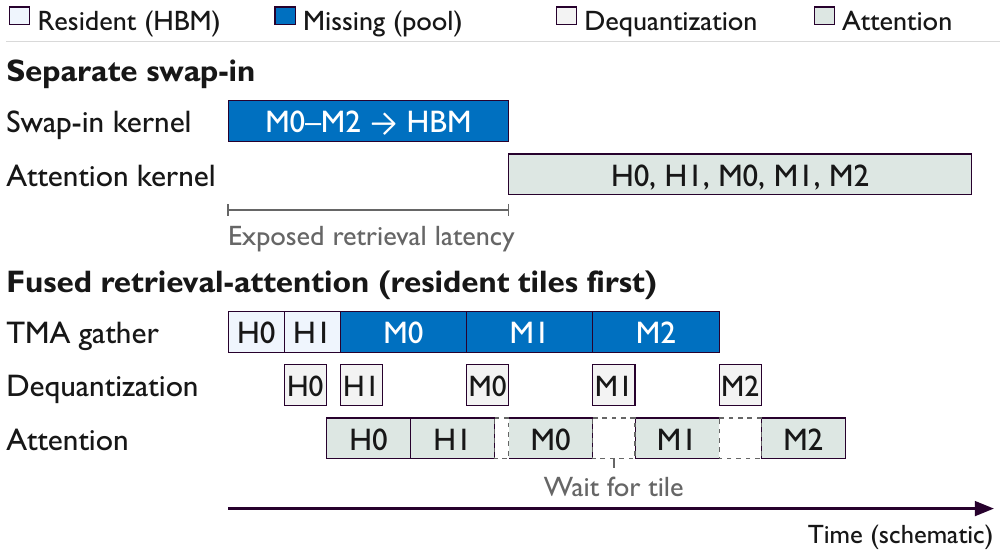}\par
    \smallskip
    (a) Separate and fused execution\par
    \medskip
    \includegraphics[width=\linewidth]{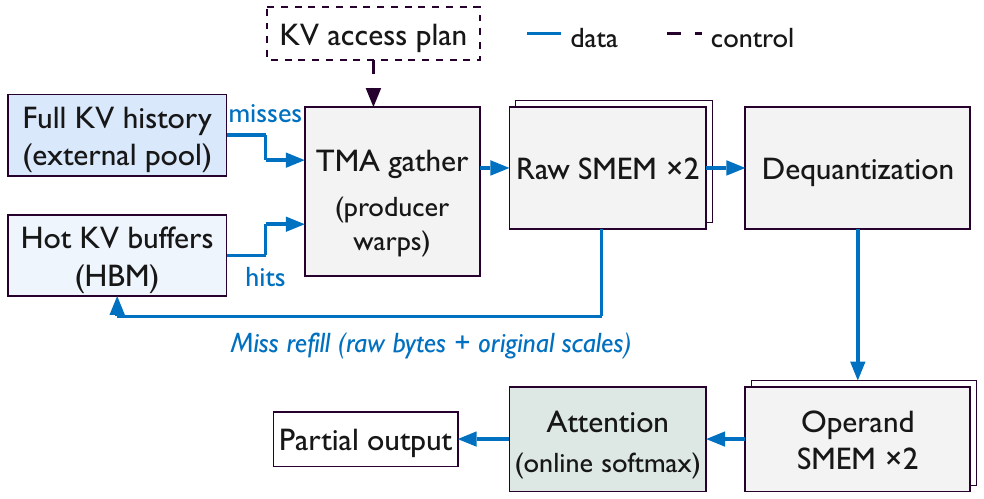}\par
    \smallskip
    (b) Buffered data flow and cache refill
    \caption{Fused KV retrieval and attention. (a) Separate swap-in and tile-level pipelining for a split containing hits and misses. (b) Shared-memory data flow and hot-buffer refill. Timelines are schematic.}
    \label{fig:design-fused-retrieval-timeline}
\end{figure}

%% file: sections/fig_code/design/k_index_prefetch.tex
\begin{figure}[t]
    \centering
    \includegraphics[width=\linewidth]{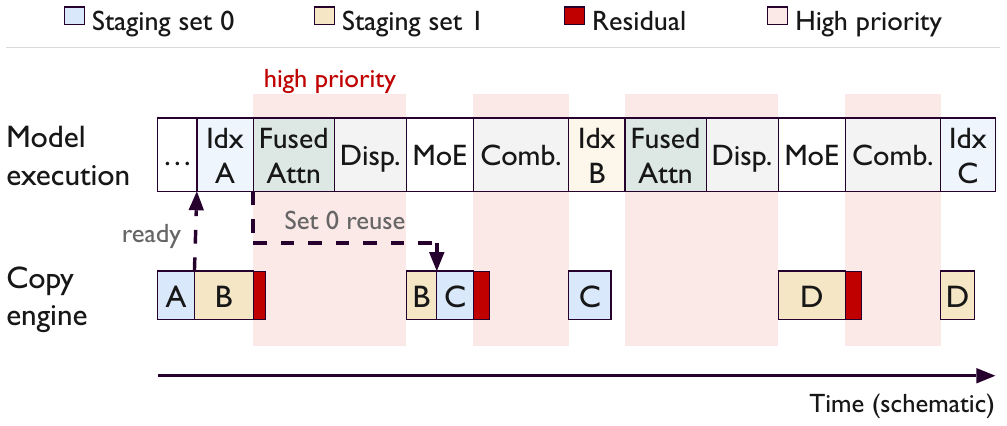}\par
    \smallskip
    (a) Cross-layer prefetching and staging\par
    \medskip
    \includegraphics[width=\linewidth]{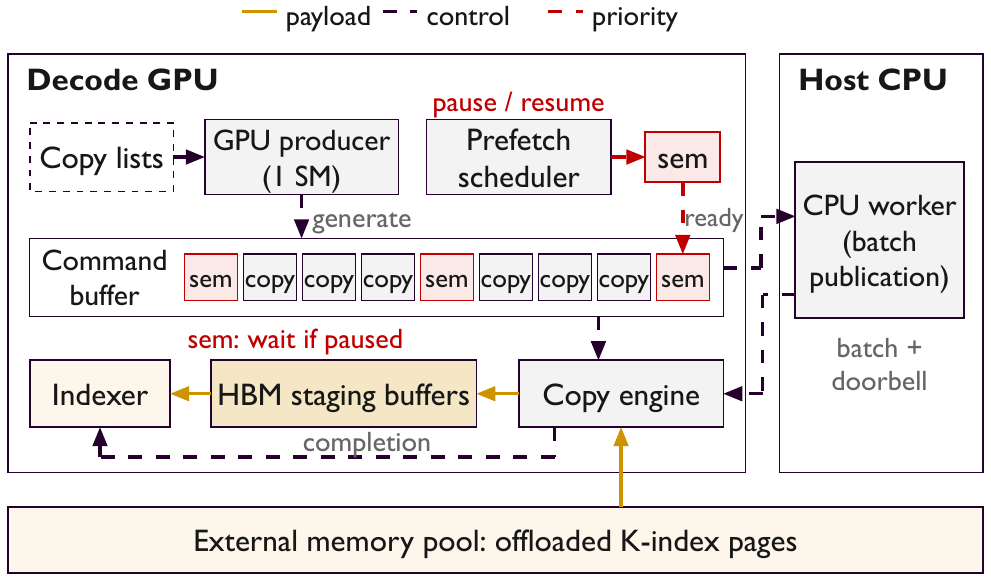}\par
    \smallskip
    (b) GPU-initiated DMA and execution control
    \caption{K-index prefetching. (a) Cross-layer staging reuse and priority scheduling. Repeated letters denote segments of one layer's prefetch; red tails show in-flight transfers completing after a pause request. Disp. and Comb. denote MoE dispatch and combine. (b) GPU command generation, CPU batch publication, and copy-engine data movement. The timeline is schematic.}
    \label{fig:design-index-prefetch}
\end{figure}

%% file: sections/evaluation.tex
\section{Evaluation}
\label{sec:evaluation}

\subsection{Experimental Setup}
\label{sec:evaluation-setup}

\paragraph{Testbed.}
We run experiments on a server with two Intel Xeon 6767P CPUs, 2~TB of host DRAM, and eight NVIDIA B300 GPUs (268~GiB HBM each).
Each GPU has 900~GB/s of unidirectional NVLink 5.0 bandwidth and a PCIe 5.0 $\times$16 link (64~GB/s per direction) for host-memory offloading.
Four GPUs run a decoding instance, and the remaining four serve as the memory nodes of the X-Mem VM~\citep{xmemvm}.
We emulate capacities beyond their 1~TB of physical memory by repeatedly mapping it into a larger virtual address space.

\input{sections/table_code/evaluation_models}

\paragraph{Models and workloads.}
We evaluate DeepSeek-V3.2 and GLM-5.3 (Table~\ref{tab:evaluation-models}) with FP8 weights, DP4 attention, and EP4 MoE.
In DP attention, each rank processes a distinct batch.
We sample requests from LongBench v2~\citep{longbenchv2} with 16K--1M input tokens and fix the output length to 128 tokens.
The default 4K-token hot KV buffer per request has an average miss rate of about 15\%.
Each decoding GPU has a 60~GiB budget for resident state, such as KV and K-index caches.

\paragraph{Baselines.}
We compare \sys{} with three baselines.
ECHO~\citep{echo}, the first open-source offloading system for sparse attention, retains a hot KV buffer on the GPU with LRU-like eviction and host-memory backing.
HiSparse~\citep{hisparse} is a production-grade sparse-attention offloading framework integrated into SGLang.
We also evaluate a naive integration, HiSparse + X-Mem, which replaces its host-memory KV backing with X-Mem.

\paragraph{Metrics.}
We compare mean time between tokens (TBT) under the same batch sizes on real traces.
We also report each system's highest measured token generation throughput per GPU and corresponding batch size.
To evaluate throughput despite physical memory constraints, we emulate workloads using a fixed 15\% miss rate and round-robin expert routing.

\subsection{End-to-End Decoding Performance}
\label{sec:evaluation-e2e}

\input{sections/fig_code/evaluation/dp_performance}

\parab{Decoding latency.}
\sys{} reduces decoding latency compared with ECHO and HiSparse (Figure~\ref{fig:evaluation-dp-latency}).
TBT reductions over ECHO are 30.0--56.0\% on DeepSeek-V3.2 and 39.6--66.6\% on GLM-5.3, compared with 7.4--35.8\% and 9.5--45.8\% over HiSparse, respectively.
At large batches, X-Mem's higher bandwidth reduces KV transfer time relative to ECHO and HiSparse, while fusion improves small-batch latency over the naive integration by overlapping retrieval and attention.

\parab{Token generation throughput.}
\sys{} achieves higher throughput than ECHO and HiSparse.
The gains over ECHO are 1.72--2.50$\times$ on DeepSeek-V3.2 and 3.14--4.98$\times$ on GLM-5.3, compared with 1.57--1.75$\times$ and 2.05--4.52$\times$ over HiSparse, respectively (Figure~\ref{fig:evaluation-dp-throughput}).
These gains primarily come from X-Mem's higher bandwidth, with its larger capacity also supporting more concurrent requests at long contexts.

\sys{} also improves throughput over the naive integration with the same X-Mem backing.
The gains are 1.10--1.34$\times$ on DeepSeek-V3.2 and 1.06--1.29$\times$ on GLM-5.3.
For example, at 128K context on GLM-5.3, the selected batch grows from 109 to 209/GPU and throughput rises from 1437.9 to 1849.3~tokens/s/GPU.
Beyond the latency savings from fusion, K-index offloading provides the main throughput gain by freeing HBM for larger batches.

\subsection{Performance Deep Dive}
\label{sec:evaluation-deep-dive}

\parab{Component ablation.}

\input{sections/table_code/evaluation_ablation}

On GLM-5.3, fusion reduces mean TBT by 15.7\% at 16K context and batch 1/GPU with offloading disabled (Table~\ref{tab:evaluation-ablation}).
K-index offloading further improves throughput by 28.6\% at 128K context, as the freed HBM allows the selected batch to grow from 109 to 209/GPU.
Appendix~A.2 provides the attention-path breakdown and cache-allocation details.

\input{sections/fig_code/evaluation/dma_efficiency}

\parab{Effectiveness of tailored I/O mechanisms.}
Our tailored I/O mechanisms improve decoding throughput during K-index offloading (Figure~\ref{fig:evaluation-dma-throughput}).
GPU-initiated DMA with priority scheduling achieves 1849.3~tokens/s/GPU, compared with 33.5 for CPU-submitted DMA and 1128.7 for SM-based offloading, a 63.8\% gain over the latter.
GPU-side command generation reduces CPU submission overhead, copy-engine transfers limit SM contention, and priority scheduling reduces interference with KV retrieval and model collectives.

GPU-initiated DMA is particularly effective for small transfers (Figure~\ref{fig:evaluation-dma-bandwidth}).
At 16~KiB, GPU-initiated DMA reaches 624.0~GB/s versus 22.4~GB/s for DMA.
The gap narrows at 1~MiB, where both approach 830~GB/s.
This convergence is consistent with submission overhead becoming less important as each message carries more data.

\input{sections/fig_code/evaluation/sensitivity_analysis}

\parab{Miss rate sensitivity.}
\sys{}'s decoding latency is less sensitive to KV misses than HiSparse's (Figure~\ref{fig:evaluation-miss-rate}).
As the miss rate rises from 0\% to 40\%, \sys{}'s mean TBT increases from 72.09 to 84.98~ms (17.9\%), while HiSparse's rises from 72.36 to 250.17~ms.
At 40\% misses, \sys{} reduces TBT by 66.0\% relative to HiSparse.
The additional KV traffic uses the higher-bandwidth X-Mem path, limiting the retrieval penalty.

\parab{Hot KV buffer size impact.}
Smaller hot KV buffers increase \sys{}'s throughput but reduce HiSparse's (Figure~\ref{fig:evaluation-hot-buffer}).
Reducing the buffer from 4K to 2K raises \sys{}'s throughput by 34.4\% at 16K context and 32.7\% at 32K, while HiSparse's throughput falls by 23.2\% and 19.5\%, respectively.
At 16K, \sys{}'s selected batch increases from 278 to 507/GPU.
Smaller buffers free HBM for larger batches, while X-Mem's higher bandwidth limits the added cost of KV retrieval.

%% file: sections/table_code/evaluation_models.tex
\begin{table}[t]
    \centering
    \caption{Model configurations.}
    \label{tab:evaluation-models}
    \small
    \setlength{\tabcolsep}{3pt}
    \renewcommand{\arraystretch}{1.08}
    \begin{tabular*}{\columnwidth}{@{\extracolsep{\fill}}l c c@{}}
        \toprule
        \textbf{Model} & \textbf{DeepSeek-V3.2} & \textbf{GLM-5.3} \\
        \midrule
        Parameters & 671B & 744B \\
        Layers & 61 & 78 \\
        K-index cross-layer reuse & No & Yes \\
        Top-$k$ tokens & 2048 & 2048 \\
        \bottomrule
    \end{tabular*}
\end{table}

%% file: sections/fig_code/evaluation/dp_performance.tex
\begin{figure*}[t]
    \centering
    \includegraphics[width=\textwidth]{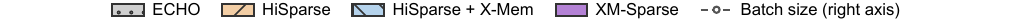}\par\nointerlineskip\vspace{1pt}
    \begin{subfigure}[t]{0.49\textwidth}
        \centering
        \begin{minipage}[t]{0.48\linewidth}
            \centering
            {\small DeepSeek-V3.2}\par\nointerlineskip\vspace{1pt}
            \includegraphics[width=\linewidth]{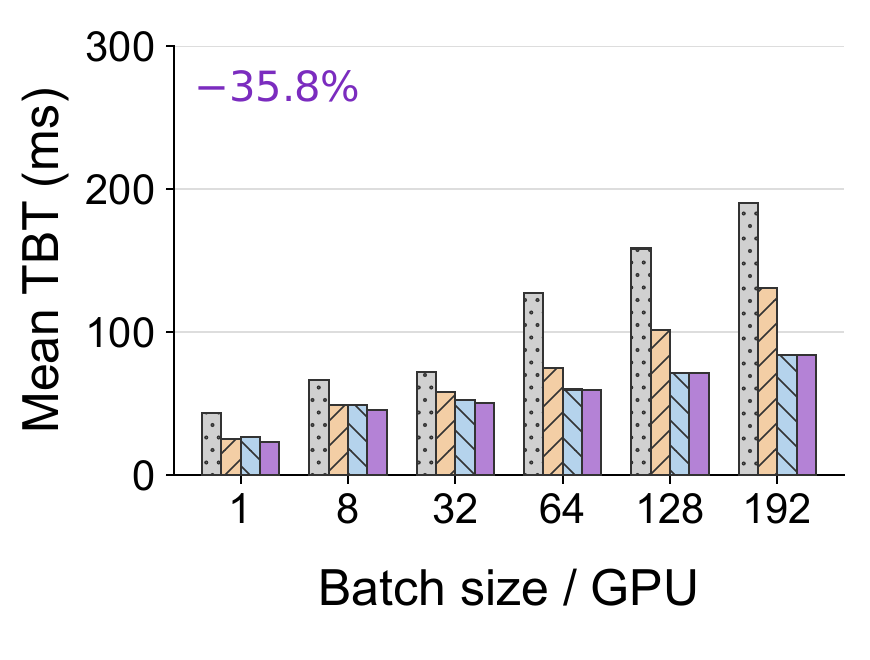}
        \end{minipage}\hfill
        \begin{minipage}[t]{0.48\linewidth}
            \centering
            {\small GLM-5.3}\par\nointerlineskip\vspace{1pt}
            \includegraphics[width=\linewidth]{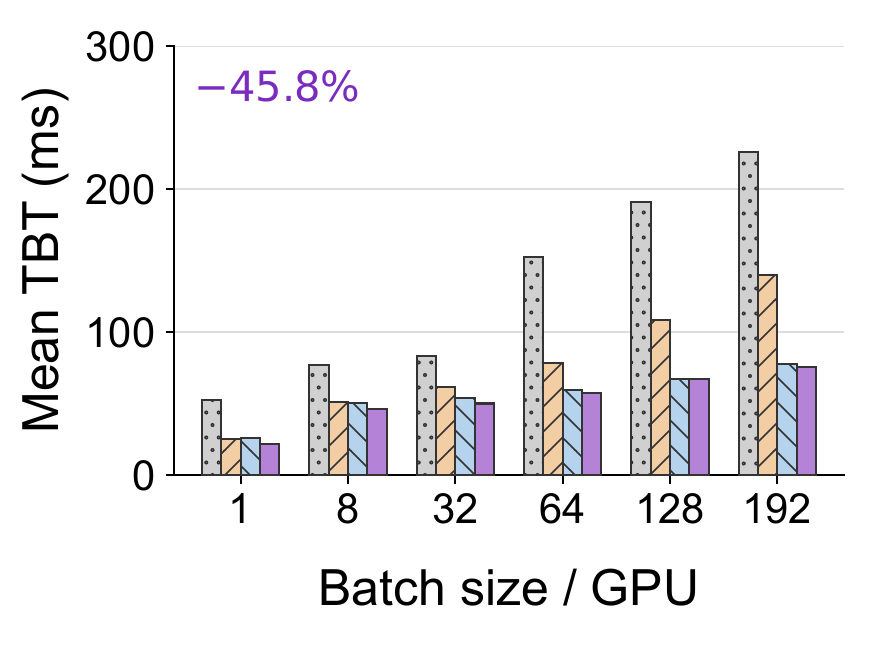}
        \end{minipage}
        \caption{Decoding latency.}
        \label{fig:evaluation-dp-latency}
    \end{subfigure}\hfill
    \begin{subfigure}[t]{0.49\textwidth}
        \centering
        \begin{minipage}[t]{0.48\linewidth}
            \centering
            {\small DeepSeek-V3.2}\par\nointerlineskip\vspace{1pt}
            \includegraphics[width=\linewidth]{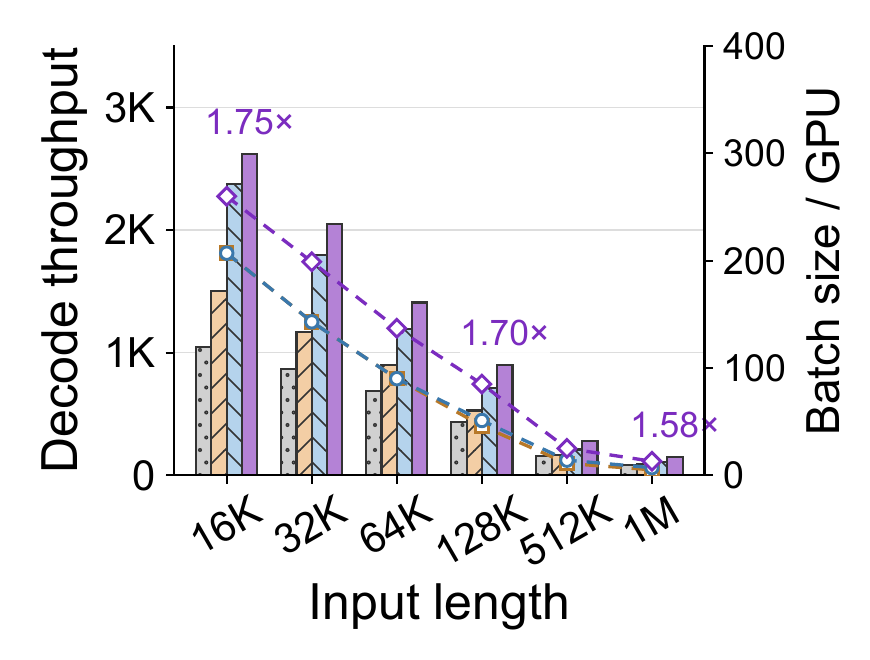}
        \end{minipage}\hfill
        \begin{minipage}[t]{0.48\linewidth}
            \centering
            {\small GLM-5.3}\par\nointerlineskip\vspace{1pt}
            \includegraphics[width=\linewidth]{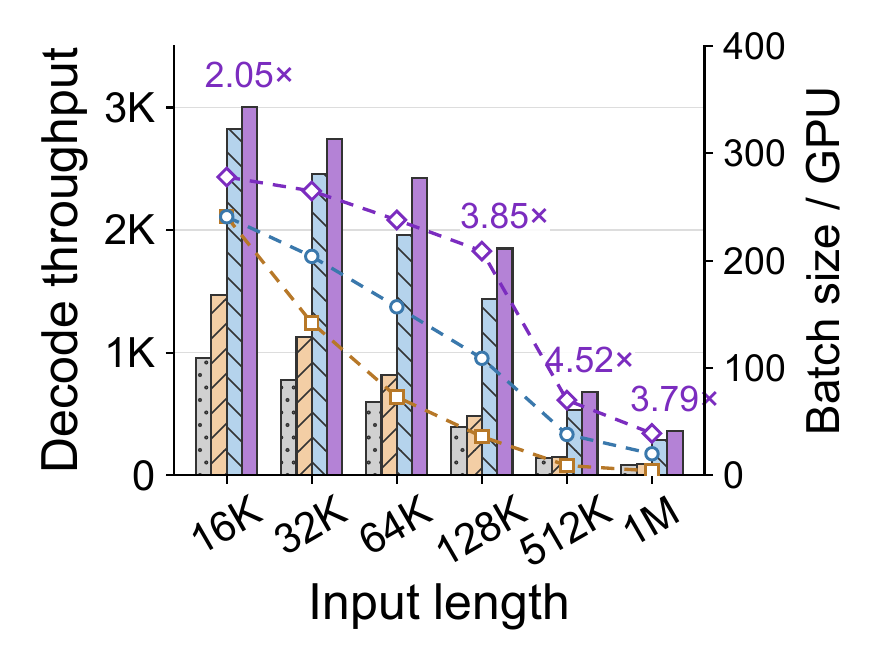}
        \end{minipage}
        \caption{Generation throughput.}
        \label{fig:evaluation-dp-throughput}
    \end{subfigure}
    \caption{End-to-end decoding performance. (a) Mean TBT at matched batches and 16K context. (b) Token generation throughput across input lengths. Dashed lines show the selected batch per GPU. Purple annotations compare \sys{} with HiSparse.}
    \label{fig:evaluation-dp-performance}
\end{figure*}

%% file: sections/table_code/evaluation_ablation.tex
\begin{table}[t]
    \centering
    \caption{Component effects on GLM-5.3.}
    \label{tab:evaluation-ablation}
    \small
    \setlength{\tabcolsep}{3pt}
    \renewcommand{\arraystretch}{1.08}
    \begin{tabular*}{\columnwidth}{@{\extracolsep{\fill}}l c c c@{}}
        \toprule
        \textbf{Optimization} &
        \textbf{Metric} &
        \textbf{Off} &
        \textbf{On (change)} \\
        \midrule
        Fused kernel &
        \makecell{Mean TBT\\(ms)} &
        25.93 & 21.86 ($\downarrow$15.7\%) \\
        K-index offload &
        \makecell{Throughput\\(tokens/s/GPU)} &
        1437.9 & 1849.3 ($\uparrow$28.6\%) \\
        \bottomrule
    \end{tabular*}
\end{table}

%% file: sections/fig_code/evaluation/dma_efficiency.tex
\begin{figure}[t]
    \centering
    \includegraphics[width=\linewidth]{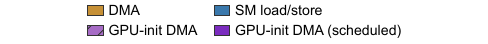}\par\smallskip
    \begin{subfigure}[t]{0.49\linewidth}
        \centering
        \includegraphics[width=\linewidth]{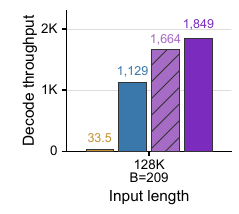}
        \caption{Decode throughput.}
        \label{fig:evaluation-dma-throughput}
    \end{subfigure}\hfill
    \begin{subfigure}[t]{0.49\linewidth}
        \centering
        \includegraphics[width=\linewidth]{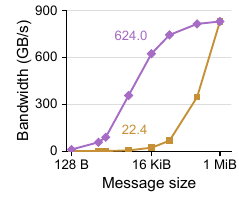}
        \caption{DMA bandwidth.}
        \label{fig:evaluation-dma-bandwidth}
    \end{subfigure}
    \caption{Offloading on GLM-5.3 at 128K and batch 209/GPU. (a) Decode throughput. Hatched bars omit scheduling. (b) Standalone bandwidth versus message size.}
    \label{fig:evaluation-dma-efficiency}
\end{figure}

%% file: sections/fig_code/evaluation/sensitivity_analysis.tex
\begin{figure}[t]
    \centering
    \includegraphics[width=\linewidth]{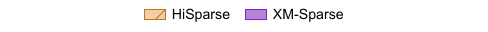}\par\smallskip
    \begin{subfigure}[t]{0.49\linewidth}
        \centering
        \includegraphics[width=\linewidth]{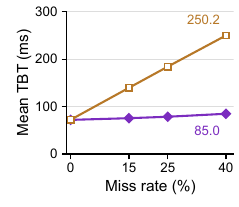}
        \caption{Miss rate.}
        \label{fig:evaluation-miss-rate}
    \end{subfigure}\hfill
    \begin{subfigure}[t]{0.49\linewidth}
        \centering
        \includegraphics[width=\linewidth]{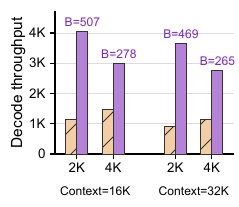}
        \caption{Hot KV buffer size.}
        \label{fig:evaluation-hot-buffer}
        \label{fig:evaluation-hot-buffer-glm}
    \end{subfigure}
    \caption{Sensitivity of GLM-5.3. (a) Mean TBT at 16K context and batch 192/GPU. (b) Throughput with 2K/25\% and 4K/15\% hot-buffer/miss configurations.}
    \label{fig:evaluation-sensitivity}
\end{figure}

%% file: sections/related.tex
\section{Related Work}
\label{sec:related}

\parab{KV-cache offloading.}
FlexGen schedules weights and KV caches across GPU, CPU, and disk~\citep{flexgen}, CachedAttention~\citep{cachedattention}, Mooncake~\citep{mooncake}, and Strata~\citep{strata} keep reusable KV caches in hierarchical memory tiers, and Pie~\citep{pie} pools CPU memory for LLM inference.
For long-context decoding, InfiniGen~\citep{infinigen} and ShadowKV~\citep{shadowkv} fetch only the KV entries predicted to be important from host memory.
DirectKV~\citep{directkv} lets GPU kernels directly access KV caches in CPU memory.

\parab{Sparse attention and memory expansion.}
Sparse-attention serving systems keep hot KV buffers and K-index cache in HBM and fetch misses from host memory over PCIe~\citep{sparseserve,hisparse,echo}, while SAC~\citep{sac} fetches them from CXL memory pool. However, CXL memory~\citep{pond,lmb} remains limited by PCIe-class bandwidth, and NVLink-C2C~\citep{nvlinkc2c2022} by capacity (Section~\ref{sec:background:memory-expansion}).
\sys{} instead targets sparse attention over scale-up memory pools, offloading both the KV and K-index caches with I/O mechanisms tailored to their access patterns.

%% file: sections/conclusion.tex
\section{Conclusion}
\label{sec:conclusion}

We presented \sys{}, a sparse-attention offloading system over scale-up memory pools.
Our analysis shows that hot KV buffers and K-index caches remain HBM bottlenecks that limit serving concurrency.
\sys{} offloads both caches with I/O mechanisms matched to their access patterns: fused TMA retrieval for KV entries and GPU-initiated DMA with priority scheduling for K-index pages.
\sys{} improves throughput by up to 4.98$\times$ over state-of-the-art baselines and 1.34$\times$ over a naive integration.

%% file: sections/appendix.tex
\raggedbottom
\section{Supplementary Material}
\label{sec:appendix}

\subsection{Bandwidth and Capacity across Context Lengths}
\label{sec:appendix-context-bottlenecks}

Using the setup in Section~3 of the main paper, we extend the bandwidth and capacity analysis to 1M context.

At 1M, external capacity becomes the tighter constraint: a 1~TB pool holds the full KV histories of only four requests per GPU (Figure~\ref{fig:appendix-context-1m}).
Expanding capacity to 24~TB permits the HBM-limited batch of 20 and yields 2.99$\times$ higher throughput over PCIe-Host.

\input{sections/fig_code/appendix/context_1m}

\subsection{Component Analysis}
\label{sec:appendix-component-analysis}

\parab{Attention-path breakdown.}
Fusion reduces latency along the attention path (Figure~\ref{fig:evaluation-fused-breakdown}).
In layer 38, the indexer-to-attention span falls from 212.2 to 163.7~$\mu$s (22.8\%), with the KV-handling and attention segments together shrinking from 83.0 to 31.4~$\mu$s.
Fusion overlaps retrieval with attention by fetching missing KV entries within the attention kernel, leaving only cache metadata updates in the preceding KV-handling stage.

\input{sections/fig_code/appendix/fused_breakdown}

\parab{HBM allocation.}
K-index offloading supports a larger batch under the same 60~GiB/GPU HBM budget (Figure~\ref{fig:evaluation-k-index-hbm}).
At 128K context, HBM allocated to K-index data and staging falls from 40.3 to 18.4~GB, while hot KV allocation grows from 24.0 to 45.9~GB.
Total cache HBM allocation remains nearly unchanged at 64.2 and 64.3~GB.
Moving K-index data to X-Mem frees HBM for more hot KV buffers, allowing the selected batch to grow from 109 to 209/GPU.

\input{sections/fig_code/appendix/k_index_hbm}

%% file: sections/fig_code/appendix/context_1m.tex
\par\addvspace{6pt}\noindent
\begin{minipage}{\columnwidth}
    \centering
    \includegraphics[width=\linewidth]{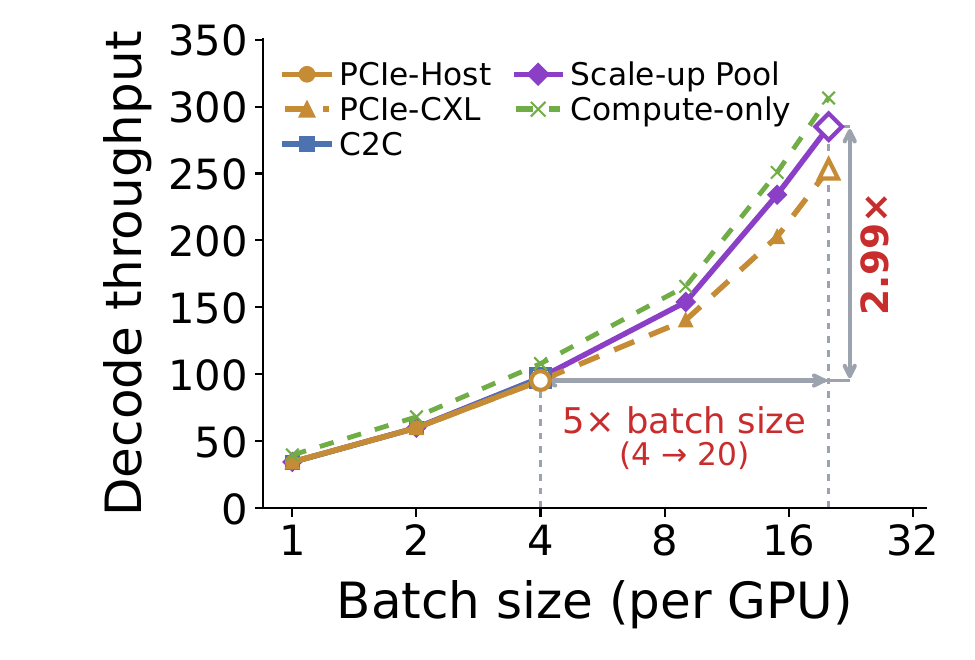}
    \captionsetup{hypcap=false}
    \captionof{figure}{Decode throughput (tokens/s/GPU) at 1M context across memory-expansion configurations.}
    \label{fig:appendix-context-1m}
\end{minipage}
\par\addvspace{6pt}

%% file: sections/fig_code/appendix/fused_breakdown.tex
\par\addvspace{6pt}\noindent
\begin{minipage}{\columnwidth}
    \centering
    \includegraphics[width=\linewidth]{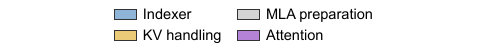}\par\smallskip
    \includegraphics[width=\linewidth]{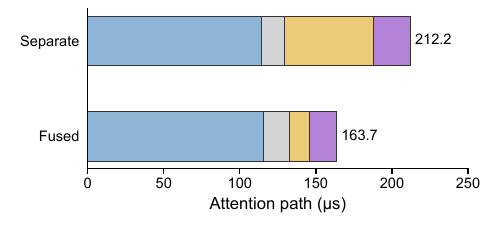}
    \captionsetup{hypcap=false}
    \captionof{figure}{Attention-path breakdown on GLM-5.3 at 16K context and batch 1/GPU. Bars show mean spans, including internal idle gaps. Attention includes split-output reduction.}
    \label{fig:evaluation-fused-breakdown}
\end{minipage}
\par\addvspace{6pt}

%% file: sections/fig_code/appendix/k_index_hbm.tex
\par\addvspace{6pt}\noindent
\begin{minipage}{\columnwidth}
    \centering
    \includegraphics[width=\linewidth]{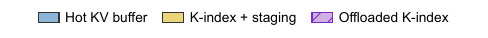}\par\smallskip
    \includegraphics[width=\linewidth]{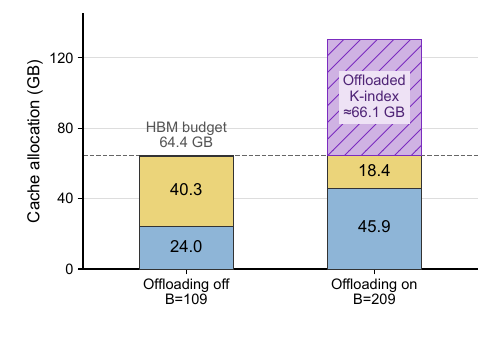}
    \captionsetup{hypcap=false}
    \captionof{figure}{HBM allocation and logical external K-index backing on GLM-5.3 at 128K context.}
    \label{fig:evaluation-k-index-hbm}
\end{minipage}
\par\addvspace{6pt}

%% file: paper.bbl
\begin{thebibliography}{37}
\providecommand{\natexlab}[1]{#1}
\providecommand{\url}[1]{\texttt{#1}}
\expandafter\ifx\csname urlstyle\endcsname\relax
  \providecommand{\doi}[1]{doi: #1}\else
  \providecommand{\doi}{doi: \begingroup \urlstyle{rm}\Url}\fi

\bibitem[Bai et~al.(2024)Bai, Tu, Zhang, Peng, Wang, Lv, Cao, Xu, Hou, Dong,
  Tang, and Li]{longbenchv2}
Bai, Y., Tu, S., Zhang, J., Peng, H., Wang, X., Lv, X., Cao, S., Xu, J., Hou,
  L., Dong, Y., Tang, J., and Li, J.
\newblock {LongBench v2}: Towards deeper understanding and reasoning on
  realistic long-context multitasks.
\newblock \emph{arXiv preprint arXiv:2412.15204}, 2024.
\newblock URL \url{https://arxiv.org/abs/2412.15204}.

\bibitem[Chen et~al.(2025)Chen, Zhang, He, Liu, Zhang, Zhou, Li, Zeng, Li,
  Qian, Li, and Li]{ess}
Chen, X., Zhang, C., He, J., Liu, W., Zhang, J., Zhou, W., Li, X., Zeng, P.,
  Li, S., Qian, Y., Li, D., and Li, Z.
\newblock {ESS}: An offload-centric latent-cache management architecture for
  {DeepSeek-V3.2-Exp}.
\newblock arXiv preprint arXiv:2512.10576, 2025.
\newblock URL \url{https://arxiv.org/abs/2512.10576}.

\bibitem[Dao et~al.(2022)Dao, Fu, Ermon, Rudra, and R{\'e}]{flashattention}
Dao, T., Fu, D.~Y., Ermon, S., Rudra, A., and R{\'e}, C.
\newblock {FlashAttention}: Fast and memory-efficient exact attention with
  {IO}-awareness.
\newblock In \emph{Advances in Neural Information Processing Systems}, 2022.

\bibitem[{Das Sharma} \& Tavallaei(2020){Das Sharma} and Tavallaei]{cxl2020}
{Das Sharma}, D. and Tavallaei, S.
\newblock {Compute Express Link} 2.0 specification now available!
\newblock {CXL} Consortium Blog, December 2020.
\newblock URL
  \url{https://computeexpresslink.org/blog/compute-express-link-2-0-specification-now-available-2374/}.

\bibitem[{DeepSeek-AI}(2025)]{deepseekv32}
{DeepSeek-AI}.
\newblock {DeepSeek-V3.2}: Pushing the frontier of open large language models.
\newblock arXiv preprint arXiv:2512.02556, 2025.
\newblock URL \url{https://arxiv.org/abs/2512.02556}.

\bibitem[Evans et~al.(2022)Evans, Andersch, Sethi, Brito, and
  Mehta]{nvlinkc2c2022}
Evans, J., Andersch, M., Sethi, V., Brito, G., and Mehta, V.
\newblock {NVIDIA Grace Hopper Superchip} architecture in-depth.
\newblock {NVIDIA} Technical Blog, November 2022.
\newblock URL
  \url{https://developer.nvidia.com/blog/nvidia-grace-hopper-superchip-architecture-in-depth/}.

\bibitem[Gao et~al.(2024)Gao, He, Sharma, Kang, Jevdjic, Deng, Yang, Yu, and
  Zuo]{cachedattention}
Gao, B., He, Z., Sharma, P., Kang, Q., Jevdjic, D., Deng, J., Yang, X., Yu, Z.,
  and Zuo, P.
\newblock Cost-efficient large language model serving for multi-turn
  conversations with {CachedAttention}.
\newblock In \emph{2024 USENIX Annual Technical Conference (USENIX ATC)}, 2024.

\bibitem[{GLM-5 Team}(2026)]{glm5}
{GLM-5 Team}.
\newblock {GLM-5}: from vibe coding to agentic engineering.
\newblock arXiv preprint arXiv:2602.15763, 2026.
\newblock URL \url{https://arxiv.org/abs/2602.15763}.

\bibitem[Lee et~al.(2024)Lee, Lee, Seo, and Sim]{infinigen}
Lee, W., Lee, J., Seo, J., and Sim, J.
\newblock {InfiniGen}: Efficient generative inference of large language models
  with dynamic {KV} cache management.
\newblock In \emph{18th USENIX Symposium on Operating Systems Design and
  Implementation (OSDI)}, 2024.

\bibitem[Li et~al.(2023)Li, Berger, Novakovic, Hsu, Ernst, Zardoshti, Shah,
  Rajadnya, Lee, Agarwal, Hill, Fontoura, and Bianchini]{pond}
Li, H., Berger, D.~S., Novakovic, S., Hsu, L., Ernst, D., Zardoshti, P., Shah,
  M., Rajadnya, S., Lee, S., Agarwal, I., Hill, M.~D., Fontoura, M., and
  Bianchini, R.
\newblock Pond: {CXL}-based memory pooling systems for cloud platforms.
\newblock In \emph{Proceedings of the 28th ACM International Conference on
  Architectural Support for Programming Languages and Operating Systems
  (ASPLOS)}, 2023.

\bibitem[Lian et~al.(2025)Lian, Tanaka, Ruwase, and Zhang]{superoffload}
Lian, X., Tanaka, M., Ruwase, O., and Zhang, M.
\newblock {SuperOffload}: Unleashing the power of large-scale {LLM} training on
  superchips.
\newblock arXiv preprint arXiv:2509.21271, 2025.
\newblock URL \url{https://arxiv.org/abs/2509.21271}.

\bibitem[Lin et~al.(2025)Lin, Tang, Yang, Wang, Tang, Tian, Stoica, Han, and
  Gao]{twilight}
Lin, C., Tang, J., Yang, S., Wang, H., Tang, T., Tian, B., Stoica, I., Han, S.,
  and Gao, M.
\newblock {Twilight}: Adaptive attention sparsity with hierarchical top-p
  pruning.
\newblock In \emph{Advances in Neural Information Processing Systems}, 2025.

\bibitem[Liu et~al.(2026)Liu, Chen, Li, Ning, Lin, Yao, Chen, Sun, Zhao, and
  Guo]{echo}
Liu, G., Chen, W., Li, C., Ning, Z., Lin, J., Yao, Y., Chen, Q., Sun, S., Zhao,
  J., and Guo, M.
\newblock {ECHO}: Efficient {KV} cache offloading with lossless prefetching for
  serving native sparse attention {LLMs}.
\newblock In \emph{20th USENIX Symposium on Operating Systems Design and
  Implementation (OSDI 26)}, pp.\  17--37, Seattle, WA, July 2026. USENIX
  Association.
\newblock ISBN 978-1-939133-55-7.
\newblock URL
  \url{https://www.usenix.org/conference/osdi26/presentation/liu-guangda}.

\bibitem[Luo \& Shen(2026)Luo and Shen]{directkv}
Luo, S. and Shen, H.
\newblock No buffer, no bottleneck: Efficient {Zero-Copy} {KV} cache offloading
  for {Long-Context} {LLMs}.
\newblock In \emph{20th USENIX Symposium on Operating Systems Design and
  Implementation (OSDI)}, pp.\  39--54, 2026.

\bibitem[Ma et~al.(2026)Ma, Ma, Li, Zha, Shang, Hu, Liu, Yang, Ma, and
  Luo]{sac}
Ma, R., Ma, T., Li, J., Zha, H., Shang, X., Hu, Q., Liu, Z., Yang, X., Ma, T.,
  and Luo, G.
\newblock {SAC}: Disaggregated {KV} cache system for sparse attention {LLMs}
  with {CXL}.
\newblock arXiv preprint arXiv:2606.19746, 2026.
\newblock URL \url{https://arxiv.org/abs/2606.19746}.

\bibitem[{Marvell}()]{marvell_cxl}
{Marvell}.
\newblock {CXL Near-Memory Compute and Expansion}.
\newblock Product Overview.
\newblock URL \url{https://www.marvell.com/products/cxl.html}.
\newblock Accessed October 28, 2026.

\bibitem[Milakov \& Gimelshein(2018)Milakov and Gimelshein]{onlinesoftmax}
Milakov, M. and Gimelshein, N.
\newblock Online normalizer calculation for softmax, 2018.
\newblock URL \url{https://arxiv.org/abs/1805.02867}.

\bibitem[{Netpreme}(2026{\natexlab{a}})]{netpreme_xm100}
{Netpreme}.
\newblock {XM-100 MPU}.
\newblock Product Overview, 2026{\natexlab{a}}.
\newblock URL \url{https://netpreme.com/products/xm-100}.
\newblock Accessed October 21, 2026.

\bibitem[{Netpreme}(2026{\natexlab{b}})]{xmemvm}
{Netpreme}.
\newblock {X-Mem} pilot virtual machine.
\newblock Early Access Program, 2026{\natexlab{b}}.
\newblock URL \url{https://netpreme.com/xmem/early-access}.
\newblock Accessed October 30, 2026.

\bibitem[{Netpreme Team}(2026{\natexlab{a}})]{netpreme2026xmem}
{Netpreme Team}.
\newblock Accelerating {SGLang HiCache} with {Netpreme X-Mem MPU}.
\newblock Netpreme Blog, July 2026{\natexlab{a}}.
\newblock URL
  \url{https://netpreme.com/blog/accelerating-sglang-hicache-with-netpreme-xmem-mpu}.

\bibitem[{Netpreme Team}(2026{\natexlab{b}})]{netpreme2026xmsparse}
{Netpreme Team}.
\newblock {XM-Sparse}: Pushing the throughput frontier of sparse attention with
  {Netpreme X-Mem}.
\newblock Netpreme Blog, October 2026{\natexlab{b}}.
\newblock URL
  \url{https://netpreme.com/blog/xm-sparse-sglang-hisparse-netpreme-xmem}.

\bibitem[{NVIDIA}({\natexlab{a}})]{dgxb300userguide}
{NVIDIA}.
\newblock \emph{{NVIDIA DGX B300 User Guide}}, {\natexlab{a}}.
\newblock URL
  \url{https://docs.nvidia.com/dgx/dgxb300-user-guide/introduction-to-dgxb300.html}.
\newblock Accessed October 28, 2026.

\bibitem[{NVIDIA}({\natexlab{b}})]{nvidia_cuda_programming}
{NVIDIA}.
\newblock \emph{{CUDA Programming Guide}: Unified Memory}, {\natexlab{b}}.
\newblock URL
  \url{https://docs.nvidia.com/cuda/cuda-programming-guide/04-special-topics/unified-memory.html}.
\newblock Accessed October 26, 2026.

\bibitem[{NVIDIA}({\natexlab{c}})]{nvidia_gb200}
{NVIDIA}.
\newblock {NVIDIA GB200 NVL72}.
\newblock Product Specifications, {\natexlab{c}}.
\newblock URL \url{https://www.nvidia.com/en-us/data-center/gb200-nvl72/}.
\newblock Accessed October 24, 2026.

\bibitem[{NVIDIA}({\natexlab{d}})]{nvidia_hopper_tuning}
{NVIDIA}.
\newblock \emph{{NVIDIA Hopper Tuning Guide}}, {\natexlab{d}}.
\newblock URL
  \url{https://docs.nvidia.com/cuda/hopper-tuning-guide/index.html}.
\newblock Accessed October 26, 2026.

\bibitem[{PCI-SIG}()]{pcisig_pcie5}
{PCI-SIG}.
\newblock {PCI Express} 5.0 frequently asked questions.
\newblock {PCI-SIG} FAQ.
\newblock URL \url{https://pcisig.com/faq}.
\newblock Accessed October 21, 2026.

\bibitem[Qin et~al.(2024)Qin, Li, He, Zhang, Wu, Zheng, and Xu]{mooncake}
Qin, R., Li, Z., He, W., Zhang, M., Wu, Y., Zheng, W., and Xu, X.
\newblock Mooncake: A {KVCache}-centric disaggregated architecture for {LLM}
  serving, 2024.
\newblock URL \url{https://arxiv.org/abs/2407.00079}.

\bibitem[Sheng et~al.(2023)Sheng, Zheng, Yuan, Li, Ryabinin, Fu, Xie, Chen,
  Barrett, Gonzalez, Liang, R{\'e}, Stoica, and Zhang]{flexgen}
Sheng, Y., Zheng, L., Yuan, B., Li, Z., Ryabinin, M., Fu, D.~Y., Xie, Z., Chen,
  B., Barrett, C., Gonzalez, J.~E., Liang, P., R{\'e}, C., Stoica, I., and
  Zhang, C.
\newblock {FlexGen}: High-throughput generative inference of large language
  models with a single {GPU}.
\newblock In \emph{Proceedings of the 40th International Conference on Machine
  Learning (ICML)}, 2023.

\bibitem[Sun et~al.(2025)Sun, Chang, Bao, Zheng, Zheng, Liu, Dong, Chi, and
  Chen]{shadowkv}
Sun, H., Chang, L.-W., Bao, W., Zheng, S., Zheng, N., Liu, X., Dong, H., Chi,
  Y., and Chen, B.
\newblock {ShadowKV}: {KV} cache in shadows for high-throughput long-context
  {LLM} inference.
\newblock In \emph{Proceedings of the 42nd International Conference on Machine
  Learning (ICML)}, 2025.

\bibitem[Tang et~al.(2024)Tang, Zhao, Zhu, Xiao, Kasikci, and Han]{quest}
Tang, J., Zhao, Y., Zhu, K., Xiao, G., Kasikci, B., and Han, S.
\newblock {Quest}: Query-aware sparsity for efficient long-context {LLM}
  inference.
\newblock In \emph{Proceedings of the 41st International Conference on Machine
  Learning}, volume 235 of \emph{Proceedings of Machine Learning Research},
  2024.

\bibitem[Wang et~al.(2024)Wang, Zhang, Tang, Chen, and Lu]{lmb}
Wang, J., Zhang, X., Tang, C., Chen, X., and Lu, T.
\newblock {LMB}: Augmenting {PCIe} devices with {CXL}-linked memory buffer.
\newblock arXiv preprint arXiv:2406.02039, 2024.
\newblock URL \url{https://arxiv.org/abs/2406.02039}.

\bibitem[Xie et~al.(2026{\natexlab{a}})Xie, Huang, Huang, Xu, Ma, and
  Kozyrakis]{hisparse}
Xie, Z., Huang, Z., Huang, T., Xu, Z., Ma, R., and Kozyrakis, C.
\newblock {HiSparse}: Scaling sparse-attention decoding with hierarchical {KV}
  cache management.
\newblock arXiv preprint arXiv:2608.07009, 2026{\natexlab{a}}.
\newblock URL \url{https://arxiv.org/abs/2608.07009}.

\bibitem[Xie et~al.(2026{\natexlab{b}})Xie, Xu, Zhao, An, Mailthody, Mahlke,
  Garland, and Kozyrakis]{strata}
Xie, Z., Xu, Z., Zhao, M., An, Y., Mailthody, V.~S., Mahlke, S., Garland, M.,
  and Kozyrakis, C.
\newblock {Strata}: Hierarchical context caching for long context language
  model serving.
\newblock In \emph{20th USENIX Symposium on Operating Systems Design and
  Implementation (OSDI 26)}, pp.\  1--16, Seattle, WA, July 2026{\natexlab{b}}.
  USENIX Association.
\newblock ISBN 978-1-939133-55-7.
\newblock URL
  \url{https://www.usenix.org/conference/osdi26/presentation/xie-zhiqiang}.

\bibitem[Xu et~al.(2024)Xu, Mao, Mo, Liu, and Stoica]{pie}
Xu, Y., Mao, Z., Mo, X., Liu, S., and Stoica, I.
\newblock Pie: Pooling {CPU} memory for {LLM} inference, 2024.
\newblock URL \url{https://arxiv.org/abs/2411.09317}.

\bibitem[Yan et~al.(2026)Yan, Karlin, and Grant]{yan2026commandstreams}
Yan, Y., Karlin, I., and Grant, R.
\newblock Revealing {NVIDIA} closed-source driver command streams for {CPU-GPU}
  runtime behavior insight.
\newblock arXiv preprint arXiv:2604.26889, 2026.
\newblock URL \url{https://arxiv.org/abs/2604.26889}.

\bibitem[Zheng et~al.(2024)Zheng, Yin, Xie, Sun, Huang, Yu, Cao, Kozyrakis,
  Stoica, Gonzalez, Barrett, and Sheng]{sglang}
Zheng, L., Yin, L., Xie, Z., Sun, C., Huang, J., Yu, C.~H., Cao, S., Kozyrakis,
  C., Stoica, I., Gonzalez, J.~E., Barrett, C., and Sheng, Y.
\newblock {SGLang}: Efficient execution of structured language model programs.
\newblock In \emph{Advances in Neural Information Processing Systems}, 2024.
\newblock URL \url{https://openreview.net/forum?id=VqkAKQibpq}.

\bibitem[Zhou et~al.(2025)Zhou, Yin, Zuo, and Cheng]{sparseserve}
Zhou, Q., Yin, P., Zuo, P., and Cheng, J.
\newblock {SparseServe}: Unlocking parallelism for dynamic sparse attention in
  long-context {LLM} serving.
\newblock arXiv preprint arXiv:2509.24626, 2025.
\newblock URL \url{https://arxiv.org/abs/2509.24626}.

\end{thebibliography}
